\documentclass[twocolumn,aps,pra]{revtex4-1}
	\usepackage[usenames,dvipsnames]{xcolor}
	\usepackage{amsmath}
	\usepackage{amsfonts}
	\usepackage{amssymb}
	\usepackage[colorlinks=true,citecolor=blue,linkcolor=red]{hyperref}
	\usepackage{graphicx}
	\usepackage{bbold}					% for \mathbb{1} as a nice unit matrix symbol
	\usepackage[makeroom]{cancel}		% crossed terms in tables
	\usepackage{multirow}				% merge cells vertically in tables
	\usepackage[normalem]{ulem}        % strike-through text
	\usepackage{array}

	\newcolumntype{x}[1]{>{\centering\let\newline\\\arraybackslash\hspace{0pt}}p{#1}}

	\DeclareMathOperator{\diag}{diag}  	% (block) diagonal matrix

	\DeclareMathAlphabet{\mathbbold}{U}{bbold}{m}{n}
	\def\abs#1{\left|{#1}\right|}      	% absolute value, \abs{x} gives |x|
	\def\bs#1{\boldsymbol{#1}}			% shorted version of "boldsymbol"
	\def\imi{\mathrm{i}}				% imaginary i
	\def\e#1{\mathrm{e}^{#1}}				% imaginary i

	\def\mcH{\mathcal{H}}					% Hamiltonian
	\def\mcT{\mathcal{T}}					% Time-reversal symmetry operator
	\def\mcK{\mathcal{K}}					% Complex conjugation operator

	\def\intg{\mathbbold{Z}}					% integers
	\def\ztwo{\mathbbold{Z}_2}					% parity group
	\def\triv{\mathbbold{0}}					% trivial group
	\def\reals{\mathbbold{R}}					% real numbers
	\def\cmplx{\mathbbold{C}}					% complex numbers
	\def\bk{{\bs{k}}}
	
	\newcounter{subeqn} %
	\makeatletter
	\@addtoreset{subeqn}{equation}
	\makeatother

\definecolor{TB}{rgb}{0,0,0} %uncomment for all black
\def\TB#1{{\color{TB}#1}}

\begin{document}
%\title{Non-trivial braiding of band-structure nodes in non-Hermitian systems}
%\title{Non-trivial braiding of Weyl points in non-Hermitian systems}
%\title{Non-trivial braiding of band nodes in non-Hermitian systems}
\title{Alice Strings in Non-Hermitian Systems}

\author{Xiao-Qi Sun$^{1,2}$}\email[]{xiaoqi@stanford.edu}
\author{Charles C. Wojcik$^3$}
\author{Shanhui Fan$^3$}
\author{Tom\'{a}\v{s} Bzdu\v{s}ek$^{2,4,5}$}\email[]{tomas.bzdusek@uzh.ch}

\affiliation{$^{1}$Stanford Center for Topological Quantum Physics, Stanford University, Stanford, CA 94305, USA}
\affiliation{$^{2}$Department of Physics, McCullough Building, Stanford University, Stanford, CA 94305, USA}
\affiliation{$^{3}$Department of Electrical Engineering, Ginzton Laboratory, Stanford University, Stanford, CA 94305, USA}
%\affiliation{$^{4}$Department of Electrical Engineering, Ginzton Laboratory, Stanford University, Stanford, CA 94305, USA}
\affiliation{$^{4}$Condensed Matter Theory Group, Paul Scherrer Institute, CH-5232 Villigen PSI, Switzerland}
\affiliation{$^{5}$Department of Physics, University of Z\"{u}rich, 8057 Z\"{u}rich, Switzerland}

\date{\today}

\begin{abstract}
%A prevailing notion in the topological band theory is that the topological charge associated with band degeneracies cannot change upon continuously tuning the Bloch Hamiltonian. Here, we show that this notion is in general incorrect in non-Hermitian systems. In particular, we present a simple three-dimensional two-band model, such that a Weyl point degeneracy flips its chiral charge after encircling an exceptional line degeneracy, upon tuning one parameter. We use the formalism of Abe homotopy to mathematically describe this phenomenon. Our work points to significant richness in the topology of non-Hermitian Hamitonians that is not shared by their Hermitian counterparts.
Alice string is a topological defect with a very peculiar feature. When a defect with a monopole charge encircles an Alice string, the monopole charge changes sign. In this work, we generalize this notion to momentum space of periodic media with loss and gain. In particular, we find that the generic band-structure node for a three-dimensional non-Hermitian crystalline system acts as an Alice string, which can flip the Chern number charge carried by Weyl points and by exceptional-line rings. 
%We discuss signatures of this topological structure for a lattice model with one tuning parameter, including non-trivial braiding of bulk band nodes, spectroscopic features of both the bulk and the surface states, and the manifest violation of the Nielsen-Ninomiya theorem.
We discuss signatures of this topological structure for a lattice model with one tuning parameter, including non-trivial braiding of bulk band nodes, and the spectroscopic features of both the bulk and the surface states. We also explore how an Alice string affects the validity of the Nielsen-Ninomiya theorem, and present a mathematical description of the braiding phenomenon.
\end{abstract}

\maketitle

\section{Introduction}
%\emph{Introduction.--} 
Alice string~\cite{Schwarz:1982,Volovik:2003} is a topological defect with a very peculiar feature. When a defect with a monopole charge encircles an Alice string, the monopole charge changes sign. In the high-energy physics context, including the Alice string leads to a highly non-trivial Alice electrodynamics~\cite{Bucher:1992}, where the charge cannot be globally defined. The mathematical structure underlying this phenomenon is the topological interaction between the homotopy invariants characterizing the Alice string and the monopole charge~\cite{Alford:1990}. Alice strings can also emerge as topological defects in various ordered phases, notably as disclination lines in nematic liquids and as vortex lines of non-chiral Bose-Einsten condensates~\cite{Leonhardt:2000,Volovik:1976,Benson:2004}, which are also described by homotopy groups~\cite{Mermin:1979}. However, although homotopic methods have been exploited to also describe nodes of energy bands in momentum space of crystalline media~\cite{Bzdusek:2017,Fang:2015,Sun:2018}, nodal lines facilitating the Alice string effect have never been reported within the extensive literature on topological band theory in Hermitian systems. 

Recently, non-Hermitian systems have attracted growing interest~\cite{Zhou:2018,Lu:2015,Chen:2016,Noh:2017,Cerjan:2018a,Wang:2017,Zhen:2015,Kozii:2017,Shen:2018,Wang:2019,Budich:2019,Yang:2019,Carlstrom:2018,Okugawa:2019,Moors:2019,Zyuzin:2018,Yoshida:2019,Zhou:2019,Zeuner:2015,Xiao:2017,Hu:2017,Poli:2015,Weimann:2016,Zhan:2017,Choi:2010,Papaj:2018,Zhong:2018,Borgnia:2019,Xu:2017,Cerjan:2018,Carlstrom:2019,Bergholtz:2019,McClarty:2019,Kawabata:2019,Lee:2018,Lee:2019,Zhou:2019,Bergholtz:2019b,Yoshida:2020} due to their rich topological structures. These systems, usually with loss and gain, are frequently modelled by non-Hermitian Hamitonians. Most of the existing studies on the topology of non-Hermitian Hamiltonians consitute perturbative corrections to Hermitian Hamiltonians. Such a treatment provides natural generalizations of concepts known from Hermitian band theory, including topological insulators and Chern numbers. In contrast, here we report an intrinsically non-perturbative property of non-Hermitian periodic media. Namely, while the Alice string phenomenon is non-existent in the stable band topology of Hermitian systems, it naturally arises in three-dimensional non-Hermitian systems. 

We find that the Alice string phenomenon in non-Hermitian systems is manifested by non-trivial braiding of band nodes in momentum space. Recall that in three-dimensional Hermitian systems, the generic band node is a Weyl point~\cite{Murakami:2007,Wan:2011}. Upon adding a non-Hermitian perturbation, a Weyl point generically inflates into an exceptional ring~\cite{Cerjan:2018,Xu:2017,Carlstrom:2019}, with the $\intg$-valued Chern number of the original Weyl point (defined on spheres) still meaningful~\cite{Shen:2018a,Yao:2018,Gong:2018,Song:2019}. Importantly, the exceptional ring is further stabilized by an additional $\intg$-valued winding number (defined on loops)~\cite{Ghatak:2019}. These two invariants correspond to the line-gap vs.~the point-gap topological classification of Ref.~\cite{Kawabata:2018}, respectively. Here, we rederive these topological charges from homotopy theory, and we show that they interact non-trivially, with exceptional line playing the role of an Alice string: the Chern number of an exceptional ring flips sign when the ring is braided around another exceptional line. As a consequence, a pair of nodes carrying the same Chern number can annihilate if they are brought together along a trajectory enclosing an exceptional line. This property makes it impossible to define the Chern number globally. %While all the other studies of topology in the non-Hermitian band are based on methods with a Hermitian analog, which are essentially perturbative. 
These results open a new viewpoint on the non-perturbative aspect of band topology in non-Hermitian systems.

The manuscript is organized as follows. First, in Sec.~\ref{sec:Chern-ambiguity} we discuss the ambiguity of Chern number in non-Hermitian models with exceptional lines. This is first motivated on theoretical grounds by considering the Riemann-sheet structure of energy bands near exceptional lines in Sec.~\ref{sec:alice-motivation}, while the subsequent Secs.~\ref{sec:model} and~\ref{sec:model2} present explicit continuum models which reveal manifestations of the Alice string effect in reciprocal braiding of band nodes. In Sec.~\ref{sec:spectra} we consider spectroscopic signatures of the Alice string effect in lattice models, which might potentially be emulated in experiments. On the one hand, we consider in Sec.~\ref{sec:spectra-bulk} the bulk band dispersion, and show that the Alice-string effect leads to bouncing of Weyl points that have been formed through a pair creation process. On the other hand, we complement this discussion in Sec.~\ref{sec:spectra-surface} by studying the properties of surface Fermi arcs of the same model. We also briefly comment here on the manifest violation of the Nielsen-Ninomiya theorem by non-Hermitian lattice Hamiltonians. Finally, in Sec.~\ref{sec:maths} we present the rigorous mathematical description of the Alice string phenomenon. We achieve this by first summarizing the construction of Abe homotopy in Sec.~\ref{sec:Abe-general}. We then explicitly apply the presented homotopic methods to non-Hermitian Bloch Hamiltonians in Sec.~\ref{sec:topo-charges}, where we rederive the Chern number and the winding number while explicitly extracting their topological interaction. In Sec.~\ref{sec:conclude} we include some concluding remarks, contrasting our results to Hermitian systems, and outlining the generalization of our arguments to non-Hermitian band insulators. Note that throughout the manuscript, by closing the energy gap we mean the formation of a band degeneracy, i.e.~a situation where two complex band energies agree both in their real and imaginary parts.
\section{Ambiguity of Chern number}\label{sec:Chern-ambiguity}
\subsection{Alice string effect of exceptional lines}\label{sec:alice-motivation}
%\subsection{Global aspects of Chern number}
%\emph{Global aspects of Chern number.--} 
Before we present an explicit model that manifests the Alice string effect, we provide an intuitive understanding of this phenomenon from general considerations. For simplicity, we first consider here a special case where the non-Hermitian effects do not inflate Weyl points into exceptional rings. The presented arguments readily generalize to exceptional rings with a Chern number.

\begin{figure}[b!]
    \centering
    \includegraphics[width=0.46\textwidth]{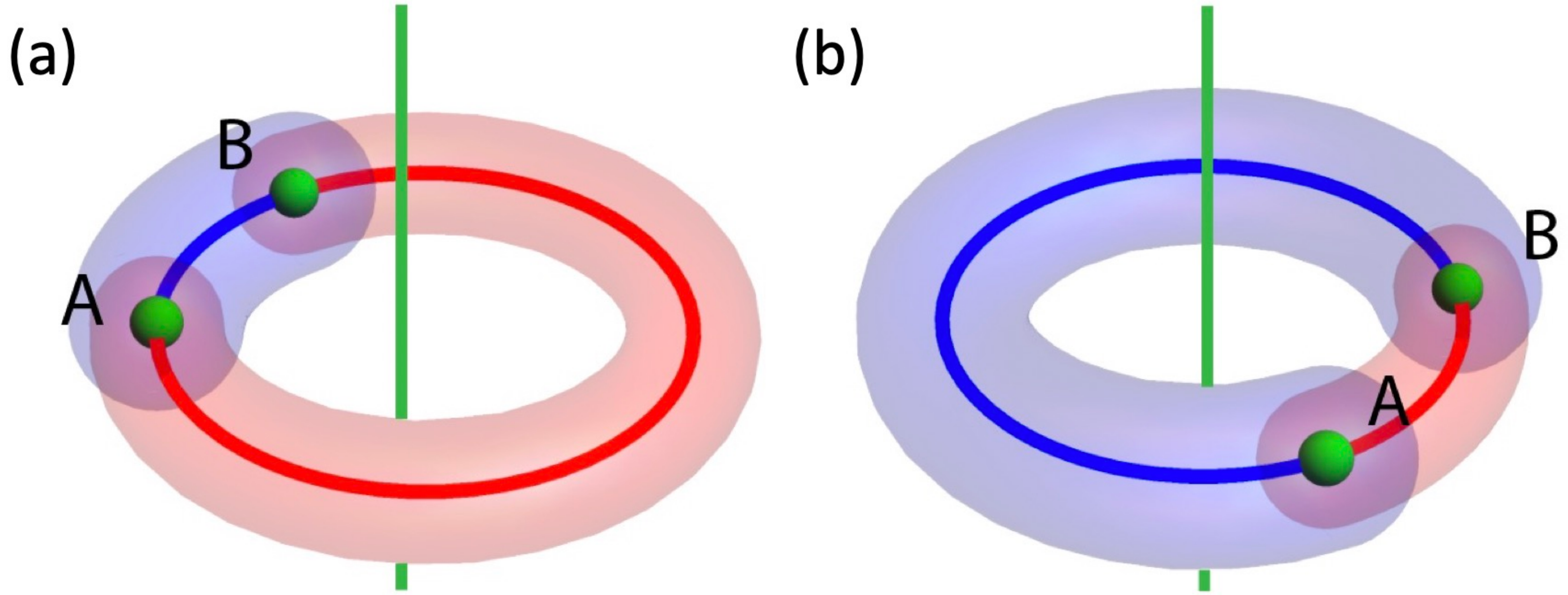}
    \caption{(a) Two Weyl points A and B (green dots) near an exceptional line (vertical green). The short blue vs.~the long red ``sausages'' represent two topologically distinct surfaces on which one can compute the total charge of the Weyl points. The total charge indicates whether the Weyl points annihilate if brought together along the blue vs.~the red trajectory contained inside the two surfaces. (b) Circumnavigating the exceptional line leads to reordering of the energy bands, which effectively flips the Chern number of the transported nodes. Therefore, if the topological charges of the Weyl points add up on the blue surface, they cancel out on the red surface.}
    \label{fig:generics}
\end{figure}

We draw in Fig.~\ref{fig:generics}(a--b) two Weyl points (green dots) near an exceptional line (vertical green line). To compute the total charge of the two Weyl points, we have two topologically distinct ways of enclosing them with a two-dimensional surface, represented by the red vs.~the blue ``sausage'' in Fig.~\ref{fig:generics}(a--b). Note that an identical figure could also illustrate a Hermitian systems with a crystalline symmetry that simultaneously enables both Weyl points and nodal lines -- in which case the green line represents a nodal line. In Hermitian systems, the two ways of computing the total Chern number of the two Weyl points give identical results. However, from topological principles, there is no guarantee for this to hold true in general, because the blue surface cannot be continuously deformed into the red surface without crossing a band node. Indeed, we find that non-Hermitian Hamiltonians provide an example where the two surfaces exhibit different total charges.

From the physics point of view, the ordering of energy eigenvalues is crucial to define Chern numbers~\cite{Sun:2018b}. However, in non-Hermitian systems the energy eigenvalues are complex-valued, hence there is no canonical choice of ordering. In fact, around an exceptional line, the band dispersion has a Riemann-sheet structure~\cite{Xu:2016,Doppler:2016,Berry:2004}, which flips the ordering of the two bands as one circumnavigates the exceptional line. Therefore, if we define an ordering of the two bands near Weyl point~A and we apply the same ordering to other momenta by continuation, we can choose one of two inequivalent paths [red vs.~blue in Fig.~\ref{fig:generics}(a--b)] to reach Weyl point~B. Since the union of these two paths encloses the exceptional line (covering one half of the Riemann sheet), the two ways of continuation give opposite ordering of the two bands near Weyl point~B, which results in opposite values of its Chern number. Especially, if the total Chern number of the Weyl points on the blue surface (enclosing the blue path) is $1\!+\!1\!=\!2$, then the Chern number on the red surface is $1\!-\!1\!=\!0$. This suggests that the Weyl points annihilate when brought together along the red trajectory, while avoiding annihilation if collided along the blue trajectory.

Note that the Riemann-sheet structure is also relevant for non-Hermitian band \emph{insulators}. If we interpret the torus (combination of the red and of the blue sausage) as the Brillouin zone (BZ) of a 2D non-Hermitian lattice system, then the non-Hermitian Hamiltonian is periodic on the torus. Nevertheless, the Chern number on the torus is ill-defined. This is because the electron wave function are not continuous function of momentum unless we consider a double cover of the BZ torus. We discuss the consequences for the classification of non-Hermitian topological band insulators in the separate work of Ref.~\cite{Wojcik:2019}.

\subsection{Continuum model for braided Weyl nodes}\label{sec:model}
\begin{figure*}[t!]
    \centering
    \includegraphics[width=0.7\textwidth]{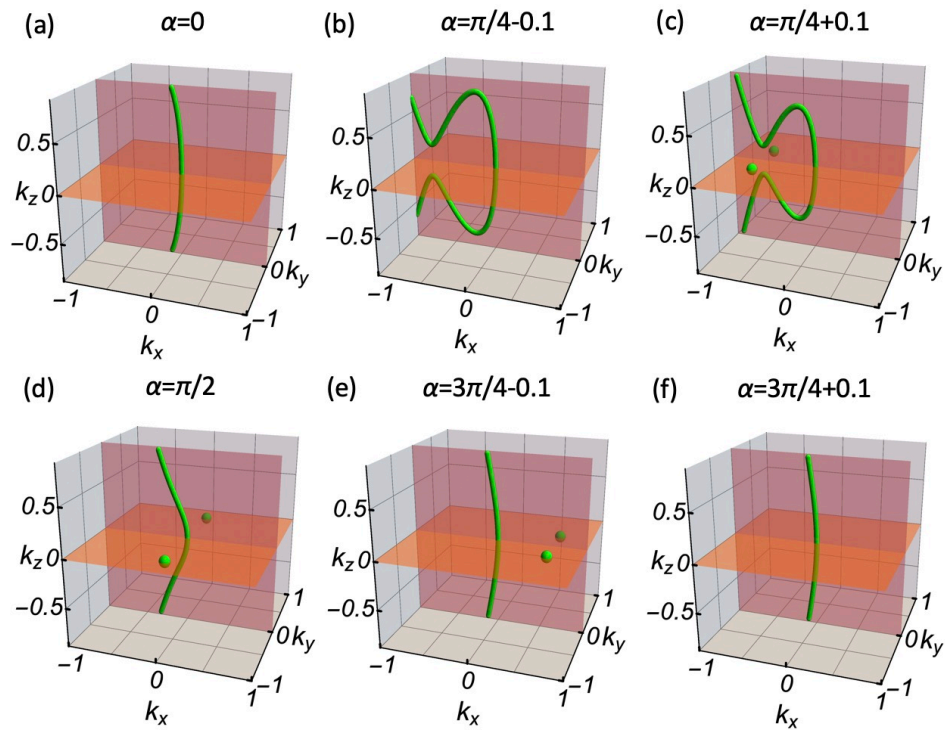}
    \caption{Band nodes of the model in Eq.~(\ref{model}) for the displayed values of $\alpha$. The orange (pink) sheet indicates the plane $k_z \!=\! 0$ ($k_y \!=\! 0)$. The band structure exhibits an exceptional line (green line) inside the pink plane. Furthermore, a pair of Weyl points (green dots) are ejected from the exceptional line at $\alpha\!=\!\pi/4$. We show in Fig.~\ref{fig:Wilson} that the ejected Weyl points locally carry the same chirality. Nevertheless, they annihilate at $\alpha\!=\!3\pi/4$ after encircling the exceptional line.}
    \label{fig:braiding}
\end{figure*}

%\subsection{Braiding of Weyl nodes around exceptional lines}
%\emph{The model.--} 
To illustrate the situation from Fig.~\ref{fig:generics} on an explicit model, we consider 
\begin{equation}
\begin{split}
\mcH(\bs{k};\alpha)&=\!\Big[2(k_{+}+e^{-\imi\alpha})(k_{-}+e^{-\imi\alpha})+1\Big]\sigma_{+}\\
&\!\!+\!\Big[2(k_{+}+e^{\imi\alpha})(k_{-}+e^{\imi\alpha})+1\Big]k_{+}\sigma_{-}+k_z \sigma_z,
\end{split} 
\label{model}
\end{equation}
where we defined 
\begin{equation}
\sigma_{\pm}=\frac{1}{2}(\sigma_{x}\pm \imi\sigma_y)\quad\textrm{and}\quad k_{\pm}=k_{x}\pm \imi k_{y} \label{eqn:Lz-op}
\end{equation}  for Pauli matrices resp.~for momentum coordinates, and where $\alpha$ is a tunable parameter. We use the k.p model in Eq.~(1) to gain an elementary intuition abound the band node braiding, and postpone the discussion of a more realistic lattice model until Sec.~III.A.
%\TB{[An explicit lattice model for the braiding appears in Sec. III.A. Here we just illustrate the situation from Fig. 1 on a simple k.p model (which we use to make certain elementary observations).]}
%A roadmap of how we constructed the Hamiltonian appears in the Supplemental Material~\cite{Supp}. (Therein, we also present an alternative \emph{lattice} model that exhibits a similar behavior.) 
The evolution of band nodes of the model in Eq.~(\ref{model}) as parameter $\alpha$ is tuned is summarized by Fig.~\ref{fig:braiding}(a--f). First, at $\alpha\!=\!0$ there is one exceptional line passing through the $k_z\!=\!0$ plane at $k_{x}\!=\!k_{y}\!=\!0$. As we increase $\alpha$ to $\pi/4$, the exceptional line ejects two Weyl points of the same chirality. As one further increases $\alpha$, the Weyl points orbit around the exceptional line in opposite directions inside the $k_z \!=\! 0$ plane, until they meet on the other side of the exceptional line at $\alpha \!=\! 3\pi/4$. Upon further increment of $\alpha$, the two Weyl points annihilate.

\begin{figure}[t!]
    \centering
    \includegraphics[width=0.5\textwidth]{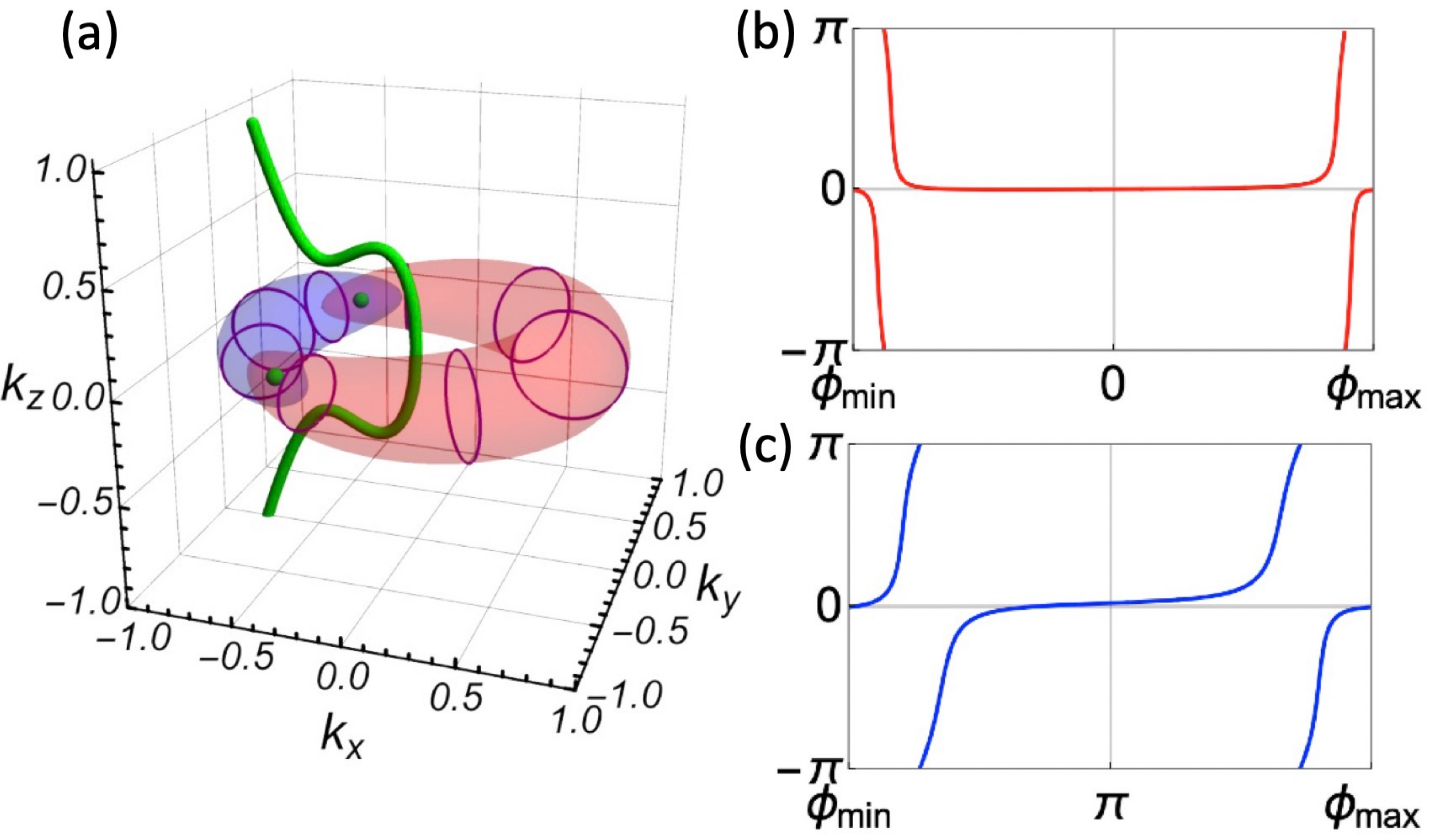}
    \caption{(a) Band structure of the model in Eq.~(\ref{model}) for $\alpha\!=\!\pi/4\!+\!0.3$ exhibits an exceptional line (green line) and a pair of Weyl points (green dots). The total charge of the two Weyl points on the blue resp.~the red surface is determined by plotting the Wilson-loop eigenvalues for paths (vertical circles) with a fixed angle $\phi \!=\! \textrm{arg}\,(k_x \!+\! \imi k_y)$, as $\phi$ sweeps along the surface. We find that the Chern number is (b) $+2$ on the blue surface, and (c) zero on the red surface. 
    }
    \label{fig:Wilson}
\end{figure}

To see that the two Weyl points ejected at $\alpha \!=\! \pi/4$ locally have the same chirality, we compute the Chern number on the blue surface displayed in Fig.~\ref{fig:Wilson}(a). This is achieved by plotting in Fig.~\ref{fig:Wilson}(b) the Wilson-loop eigenvalues~\footnote{The definition of the Wilson loop operator and the corresponding Berry phase depends on the normalization of the states in non-Hermitian systems. The convention of this paper corresponds to the left-right Berry curvature in Ref.~\cite{Shen:2018a}} for paths that sweep along the surface~\cite{Gresch:2017}. The observed winding indicates that the total Chern number on the blue surface containing the two Weyl points just after their conception is $+2$. Meanwhile, since the two Weyl points annihilate for $\alpha \!=\! 3\pi/4$ on the other side of the exceptional line, the total Chern number on the red surface in Fig.~\ref{fig:Wilson}(a) must be zero.  This is confirmed by plotting the corresponding Wilson-loop eigenvalues in Fig.~\ref{fig:Wilson}(c). We conclude that the Chern number of Weyl points in non-Hermitian systems exhibits an ambiguity: we are able to flip the Chern number of a Weyl point by moving it around an exceptional line, exactly as anticipated in the Alice string phenomenon.

In general, the stability of a line defect (here exceptional lines) is captured by a topological charged derived from first homotopy group of a classifying space of Hamiltonians, while the stability of a monopole charge (Weyl points in the above argument) follows a topological charge derived from second homotopy group~\cite{Bzdusek:2017}. The observed interaction between the two invariants, corresponding to the Alice string effect~\cite{Schwarz:1982}, can be described by Abe homotopy~\cite{Abe:1940}. Abe's approach encompasses both the first and the second homotopy group within a single construction on a cylinder, and therefore it allows for a natural description of their interaction. We provide a brief summary of this mathematical theory in Sec.~\ref{sec:Abe-general}, and we discuss the explicit application of these methods to non-Hermitian Bloch Hamiltonians in Sec.~\ref{sec:topo-charges} near the end of the manuscript.

\subsection{Braiding through an inflated Weyl node}\label{sec:model2}

Before we discuss in next sections the spectroscopic signatures and the underlying mathematical description of the Alice string phenomenon, we consider here one more instance of non-trivially braided band nodes. In this example, the Chern number of a nodal-line ring becomes ambiguous due to the presence of Weyl points elsewhere in $\bs{k}$-space. This is a natural counterpart to the ambiguous chirality of Weyl points that are moving around an exceptional line, demonstrated by the model in Eq.~(\ref{model}). %Both phenomena indicate that Chern number cannot be globally defined in non-Hermitian systems.

\begin{figure*}[t!]
    \centering
    \includegraphics[width=0.7\textwidth]{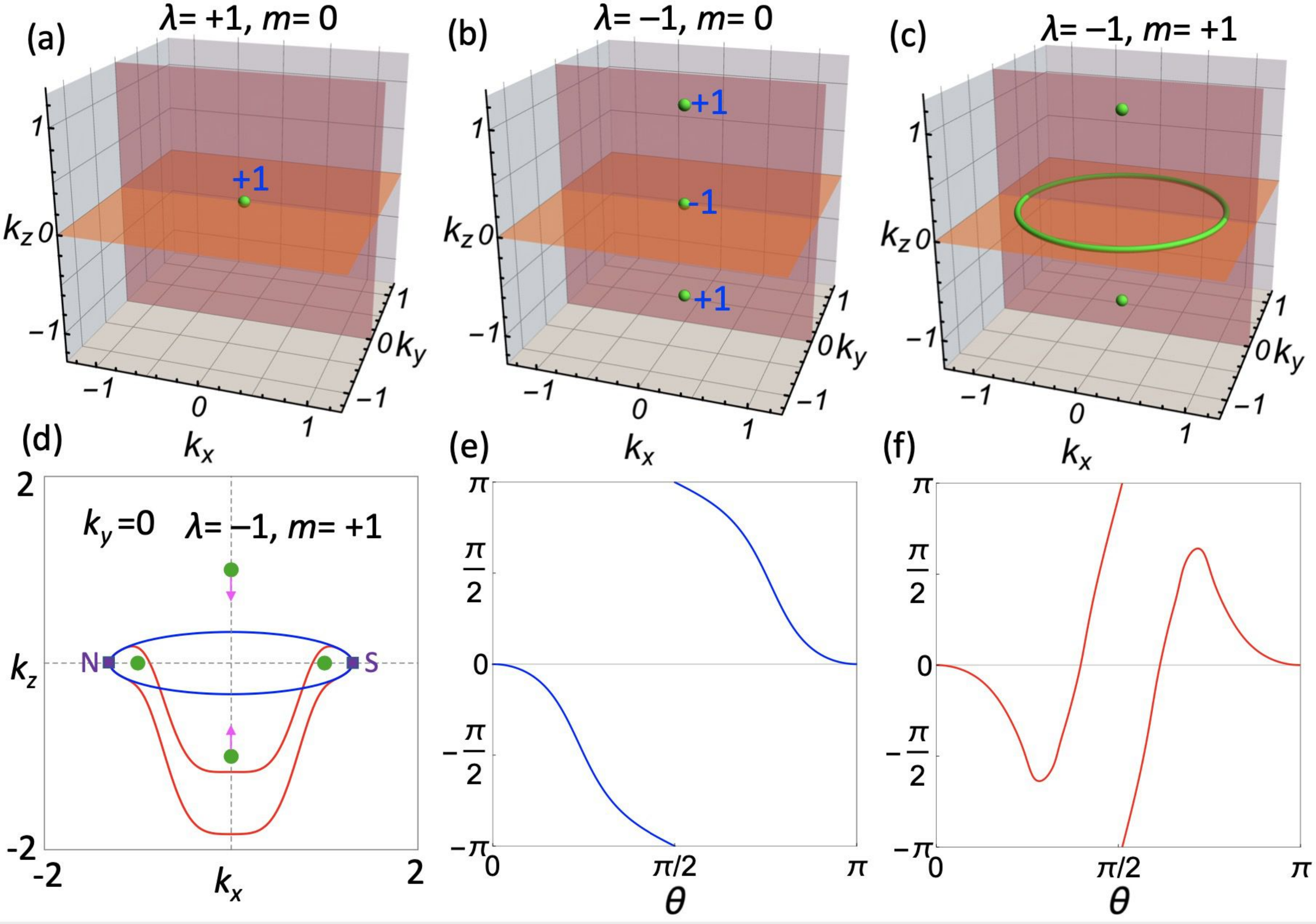}
    \caption{(a--c) Band nodes (green) of the Hamiltonian in Eq.~(\ref{eqn:model-2}) for the indicated values of parameters $\lambda$ and $m$. For $m\!=\!0$, the model is Hermitian, and one can assign a unique chirality to all the Weyl points (the $+$ vs.~$-$ signs displayed with blue font). For $m\!\neq\! 0$, Chern number of a node becomes ambiguous and depends on the choice of the enclosing surface. The non-Hermitian perturbation is set up such that it inflates into an exceptional ring only the Weyl point at $k_z\!=\!0$. (d) The band nodes of panel (c) inside the pink sheet $k_y\!=\!0$. To study the monopole charge of the exceptional ring, we consider surfaces of revolution obtained by rotating around the $k_z$-axis the blue loop [$k_x = 1.3\cos\phi$ and $k_z = \tfrac{1}{3}\sin\phi$ for $\phi\in[0,2\pi)$] resp.~the red loop [$k_z$ further lowered by $\tfrac{3}{2}(\cos^4\phi -1)^8$]. We describe the obtained surfaces as an ``ellipsoid'' resp. a ``bowl''. The Weyl point at $k_z \!<\!0$ makes it impossible to deform the ellipsoid into the the bowl without closing the energy gap, therefore the two surfaces may exhibit different Chern number. (e,f) Wilson-loop flow for the ellipsoid resp.~the bowl surface. In both cases, the Wilson operators are computed along paths of constant $k_x$, and the parameter $\theta$ grows from $0$ at point $\mathrm{N}$ to $\pi$ at point $\mathrm{S}$, indicated for both surfaces in panel (d). The branch of the Riemann sheet is chosen such that both surfaces consider the same Bloch state at points $\mathrm{N}$ and $\mathrm{S}$, where both eigenvalues of the Hamiltonian remain real for all considered values of the model parameters.
    }
    \label{fig:model-2}
\end{figure*}

To make our point, we consider the model
\begin{equation}
\mcH(\bs{k};\lambda,m) = k_x \sigma_x + k_y \sigma_y + (k_z + \imi m)(k_z^2 + \lambda)\sigma_z\label{eqn:model-2}
\end{equation}
where $\lambda$ and $m$ are tunable parameters.~\footnote{One can change this $\bs{k}\cdot\bs{p}$ model into a lattice model through the usual substitution $k_i \mapsto \sin k_i$ and $k_i^2\mapsto 2(1-\cos k_i)$, which may however result in many copies of the band nodes at other high-symmetry momenta. However, the non-Hermiticity of the system allows for an alternative substitution $k_i \mapsto 2\sin \tfrac{k_i}{2}\exp{(\imi k_i/2)}$ and $k_i^2\mapsto 2(1-\cos k_i)$ that results in fewer band nodes elsewhere in the Brillouin zone.} The Hamiltonian in Eq.~(\ref{eqn:model-2}) is invariant under antiuniary symmetry $C_{2z}\mcT: k_z \mapsto -k_z$ (composition of time-reversal with $\pi$-rotation around $z$-axis), represented by $\sigma_x \mcK$. We study how the nodal structure of the Hamiltonian evolves as parameters $\lambda$ and $m$ are tuned along the following path.
\begin{itemize}
\item First, we set $\lambda \!=\! +1$ and $m\!=\!0$. This corresponds to a Hermitian Hamiltonian with a single Weyl point with chirality $\chi\!=\!+1$ at $\bs{k} \!=\! \bs{0}$, as shown in Fig.~\ref{fig:model-2}(a).
\item Keeping $m\!=\!0$, we lower the other parameter to $\lambda \!=\! -1$. This preserves the Hermiticity of the model, but a phase transition occurs at $\lambda\!=\!0$, in which the Weyl point at $\bs{k} \!=\! 0$ flips chirality to $\chi \!=\! -1$ while ejecting two Weyl points with chirality $\chi\!=\!+1$~\cite{Sun:2018}. These additional Weyl points are related to each other by $C_{2z}\mcT$, and at $\lambda \!=\! -1$ they are located at $\bs{k}\!=\!(0,0,\pm 1)$, see Fig.~\ref{fig:model-2}(b).
\item Then, while preserving $\lambda \!=\! -1$, we turn on a non-Hermitian perturbation by setting $m\!=\!+1$. This keeps the two outer Weyl points fixed, but inflates the Weyl point at $\bs{k}\!=\!\bs{0}$ into an exceptional ring at $k_z \!=\! 0$ and $k_x^2 + k_y^2 \!=\! 1$, as plotted in Fig.~\ref{fig:model-2}(c).
\end{itemize}
We already learned in Sec.~\ref{sec:model} that the chirality of Weyl points becomes ambiguous in non-Hermitian models with exceptional lines. In the present case, the two outer Weyl points in Fig.~\ref{fig:model-2}(c) can be brought together and \emph{annihilate at $\bs{k}\!=\!\bs{0}$ inside the exceptional ring} if moved along the magenta arrows in Fig.~\ref{fig:model-2}(d), even though in the original Hermitian model they have carried the same (positive) chirality, cf.~Fig.~\ref{fig:model-2}(b). The exceptional ring can finally be contracted to a Weyl point again while making the resulting Hamiltonian Hermitian (not illustrated in Fig.~\ref{fig:model-2}). These last two steps require breaking of the $C_{2z}\mcT$ symmetry and go beyond the simple Hamiltonian provided by Eq.~(\ref{eqn:model-2}), but such a process is in principle possible. We will discuss the details of this process in Appendix~\ref{app:annihilation}.

The evolution of band nodes outlined in the previous paragraph seems contradictory. Note that we have departed from a Hermitian model with a single Weyl point with \emph{positive} chirality in Fig.~\ref{fig:model-2}(a), and we end up with a Hermitian model that contains a single Weyl point -- namely one obtained from the \emph{negative} chirality Weyl point in Fig.~\ref{fig:model-2}(b). The resolution to this paradox lies in the words ``\emph{obtained from}''. While the (negative) chirality of the central Weyl point in Fig.~\ref{fig:model-2}(b) is carried by the blue ``ellipsoid'' surface in Fig.~\ref{fig:model-2}(d), the (actually positive) chirality of the Weyl point in the final Hermitian model is carried by the red ``bowl'' surface in Fig.~\ref{fig:model-2}(d). This is because annihilation of the two outer Weyl points inside the exceptional ring implies closing of the energy gap on the ellipsoid but not on the bowl surface. Although both surfaces enclose the same exceptional ring, the presence of the intermediate Weyl point makes it impossible to deform one surface onto the other without closing the energy gap. As a consequence, the two surfaces may carry a different Chern number, akin to the situation in Fig.~\ref{fig:Wilson}. Our suspicion is indeed confirmed by the Wilson-loop data for the ellipsoid and for the bowl surface, plotted respectively in Fig.~\ref{fig:model-2}(e) and~(f). We thus observe that moving the Weyl point through the exceptional ring reverses the monopole charge of the ring. More generally, one can show that moving a monopole charge $n$ through an exceptional ring changes the monopole charge of the ring by $- 2 n$.

The observed evolution of band nodes and of their computed topological charges for the models given by Eqs.~(\ref{model}) and~(\ref{eqn:model-2}) convincingly demonstrate that Chern number becomes ambiguous in non-Hermitian models, confirming that exceptional line acts as an Alice string. While in Hermitian models the Chern number on a surface $\partial \mathcal{D}$ can be obtained by integrating inside region $\mathcal{D}$ a local density, namely the divergence of Berry curvature $\frac{1}{2\pi}\bs{\nabla}\cdot\bs{F} = \sum_i \chi_i\delta(\bs{k}-\bs{k}_i)$, no such a local-density formulation exists for Chern number in non-Hermitian systems. Following the terminology of Ref.~\cite{Alford:1990}, Chern number in non-Hermitian systems becomes a ``Cheshire charge''.

\section{Spectroscopic signatures}\label{sec:spectra}
\subsection{Bulk signatures}\label{sec:spectra-bulk}
%\section{Bulk spectroscopic signatures}
%\emph{Spectroscopic signatures.--} 

\begin{figure*}[t!]
    \centering
    \includegraphics[width=0.9\textwidth]{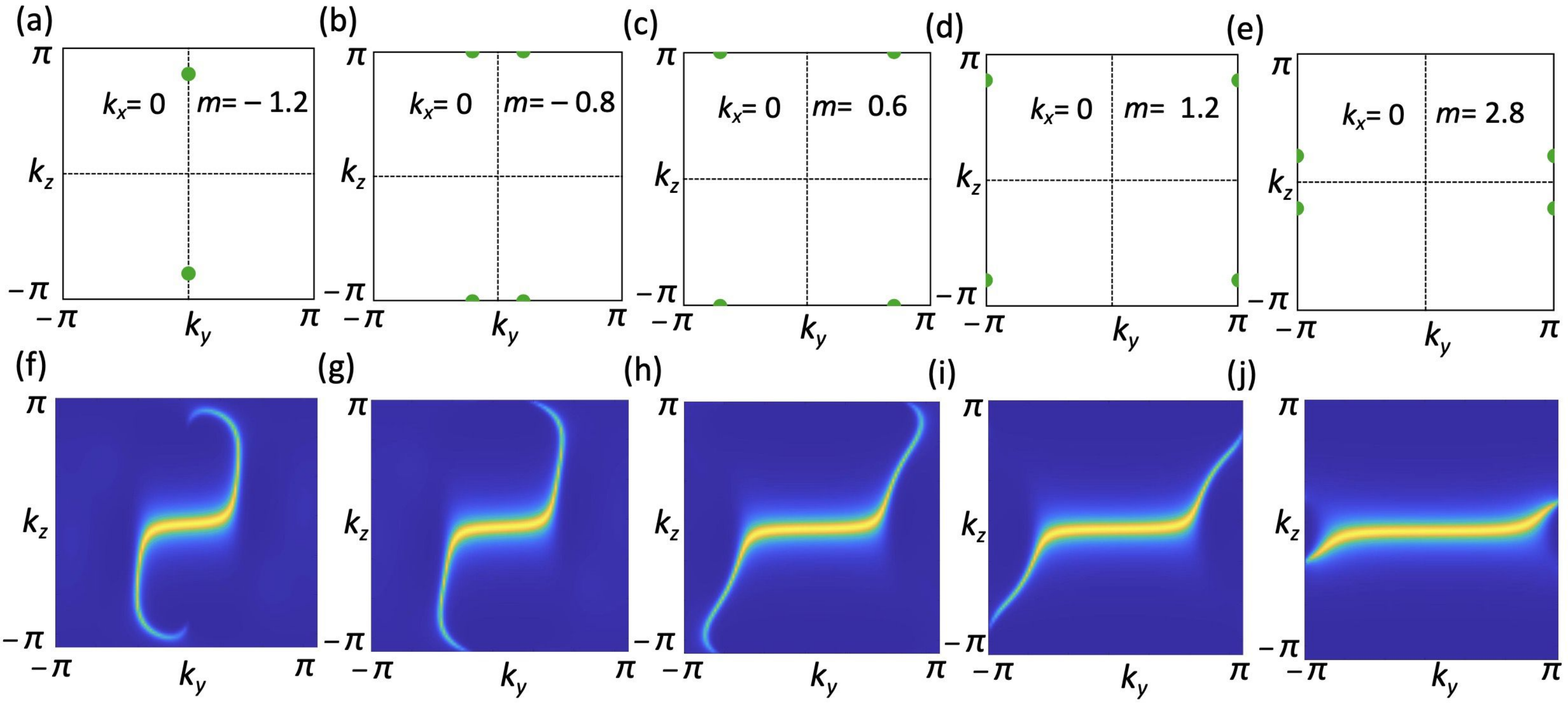}
    \caption{(a--e) Position of Weyl points (green dots) of the model in Eq.~\eqref{lattice-model} inside the $k_x\!=\!0$ plane for the indicated values of parameter $m$. At $m\!=\!-3$, a pair of Weyl points of opposite chirality are created at $(k_y,k_z)=(0,0)$, and depart along the $k_z$ axis. The Weyl points collide and bounce for $m\!=\!-1$ at $(k_y,k_z)=(0,\pi)$ after circumnavigating the $k_z$-circle of the Brillouin zone (BZ) torus, and then again 
    %. Another collision and bounce of the Weyl points occurs 
    for $m\!=\!+1$ at $(k_y,k_z)=(\pi,\pi)$. The Weyl points do not annihilate at these two collisions because of the Alice string enclosed by the $k_z$-circle of the BZ torus. They Weyl points finally annihilate for $m\!=\!+3$ at $(k_y,k_z)=(\pi,\pi)$ after the effect of the Alice string is undone by moving along the $k_z$-direction for the second time. (f--j) Zero-energy surface Fermi arcs of the same model for the corresponding values of $m$ assuming a system termination in the $x$-direction. The details of the computation and plotting are explained in the second paragraph of Sec.~\ref{sec:spectra-surface}.}
    \label{fig:conversion}
\end{figure*}

Due to the high controllability and tunability, there have been experimental studies of both Weyl points and exceptional lines in photonic systems~\cite{Lu:2015,Chen:2016,Noh:2017,Cerjan:2018a}. In fact, the existing experimental techniques readily allow for inferring the topological charges of both exceptional nodes~\cite{Zhou:2018} and Weyl points~\cite{Wang:2017} in such systems. However, these techniques are rather indirect. Here we present one way of demonstrating the Alice string effect with a simple signature, namely by observing the spectral evolution upon continuously changing the Hamiltonian parameters. 

As a proof-of-principle thought experiment, we consider the non-Hermitian Hamiltonian on a lattice
\begin{subequations}\label{eqn:lattice-model}
\begin{equation}
\mcH(\bs{k};m) = \e{\imi k_z /2 }H_0(\bs{k};m)
\end{equation}
with Hermitian matrix
\begin{equation}
\begin{split}
H_0(\bk;m)&=\e{+\imi \frac{k_x-k_z}{2}}[a(\bk;m)+\imi b(\bk)]\sigma_{+}+\\
&\hspace{-0.6cm}\e{-\imi \frac{k_x-k_z}{2}}[a(\bk;m)-\imi b(\bk)]\sigma_{-}+ h_{z}(\bk)\sigma_z
\end{split}
\label{lattice-model}
\end{equation}
and with %the explicit form of the 
real functions %$a(\bk;m)$, $b(\bk)$ and $h_z(\bk)$ are
\begin{equation}
\begin{split}
&a(\bk;m)=\left[\frac{m}{2}+\cos k_y\left(1+ \frac{\cos{k_z}}{2}\right)\!\right]\cos \frac{k_x}{2},\\
&\;b(\bk)=\sin \frac{k_x}{2}\quad\,\textrm{and} \quad\, h_{z}(\bk)=\sin k_y  \cos \frac{k_z}{2}.
\end{split}
\end{equation}
\end{subequations}
The Hamiltonian $\mcH(\bs{k};m)$ defined by Eqs.~(\ref{eqn:lattice-model}) has period $2\pi$ in all momentum components $(k_x,k_y,k_z)$, and it can be realized on a cubic %square
lattice. In contrast, the matrix $H_0$ does not respect the $2\pi$-periodicity in the $k_z$-direction, but instead it exhibits a doubled $4\pi$ period. The eigenvalues of the non-Hermitian Hamiltonian in Eq.~(\ref{eqn:lattice-model}) are
\begin{equation}
E_{\pm}(\bk;m)=\pm \e{\imi\frac{k_z}{2}}\sqrt{a(\bk;m)^2+b(\bk)^2+h_z(\bk)^2}.
\end{equation}
As a function of $k_z$, the two energy eigenvalues exchange from $k_z=0$ to $k_z=2\pi$, reminiscent of the Riemann-sheet structure near an exceptional line. In fact, the Hamiltonian carries a non-trivial value of the winding number defined from the first homotopy group. Conceptually, the exceptional line is enclosed by the $k_z$-circle of the 3D Brillouin zone (BZ) torus. 

Importantly, the model in Eq.~(\ref{eqn:lattice-model}) supports a pair of Weyl points for $\abs{m} \! < \! 3$, which move non-trivially inside the $k_x\!=\!0$ plane as illustrated in Fig.~\ref{fig:conversion}(a--e). On this plane, upon increasing the value of $m$, we first create a pair of Weyl points with opposite chirality at $(k_y,k_z)=(0,0)$, which then depart in opposite directions along the $k_z$-axis. The Weyl points meet at $(k_y,k_z)=(0,\pi)$ for $m=-1$, after circumnavigating the $k_z$-circle of the BZ torus. Since the trajectory of the two Weyl points has enclosed one branch of the Riemann sheet (the ``exceptional line" wrapped by the BZ torus), the relative chirality of the Weyl points has been flipped and they cannot annihilate. Instead, they bounce in the opposite directions along the $k_y$-axis, as illustrated in Fig.~\ref{fig:conversion}(a,b). 
%Further calculation of the Wilson loop flow can verify that the two Weyl points carries same Chern number at the bouncing point. 
As an experimental signature, we can identify the path-dependent capability of Weyl points to annihilate (i.e.~to open a spectral gap) solely from spectrum measurements upon changing the parameter $m$, i.e.~without accessing the topological invariant carried by the band nodes. 

Further increasing the parameter to $m\!=\!+1$ leads to another collision of the Weyl points at $(k_y,k_z) = (\pi,\pi)$, after moving around the $k_y$-circle of the BZ torus. Since this direction is not associated with a non-trivial winding number, the relative chirality of the Weyl points is not modified since their previous encounter, and they bounce in opposite directions along the $k_z$-axis, see Fig.~\ref{fig:conversion}(c,d). Finally, at $m\!=\!+3$, the Weyl meet at $(k_y,k_z) = (\pi,0)$ after moving around the non-trivial $k_z$-direction of the BZ torus for the second time. This flips their relative chirality, allowing them to annihiliate, leaving behind a band insulator for $m\!>\!3$. Since the Weyl points exhibited a net motion along the $k_y$-direction, the resulting model at $m\!>\!3$ is a weak Chern insulator.

We also point out the manifest violation of the Nielsen-Ninomiya (NN) no-go theorem by the lattice model in Eq.~(\ref{eqn:lattice-model}). The NN theorem~\cite{Nielsen:1981} states that the total chirality of Weyl nodes exhibited by a lattice model with local hoppings has to be zero. In constrast, our model for $m\!=\!-1$ (for $m\!=\!+1$) exhibits one double Weyl point at $\bs{k}=(0,0,\pi)$ [at $\bs{k}=(0,\pi,\pi)$] with no counterpart elsewhere in the BZ. The violation is possible, because the proof of the NN theorem assumes that one can uniquely order the energy bands from lowest to highest. Such an ordering becomes ambiguous in non-Hermitian due to the complex-valued nature of the energy bands, especially if the winding number (the first-homotopy invariant) becomes non-zero on some closed path. The observed violation of the NN theorem is analogous to the loophole reported for Floquet systems (the periodic nature of quasienergy)~\cite{Sun:2018b,Higashikawa:2019}, and is fundamentally different from the strategy of Ref.~\cite{Yu:2019} for static Hermitian models (inflating one of the Weyl points into a nodal surface covering the BZ boundary).

%\TB{\subsection{Surface signatures}}
%\TB{\section{Surface spectroscopic signatures}}

%\TB{Can we say something about the surface Fermi arcs? Yes, but include surface perturbation $\lambda \e{\imi k_z/2}\sin(\tfrac{k_z}{2}) \sigma_z$ with $\lambda = ?$. [XiaoQi, do I remember this correctly?] This is to explicitly break the antiunitary symmetry $C_{2y}\mcT: (k_x,k_y,k_z) \mapsto (-k_x,k_y,k_z)$ symmetry represented by $\sigma\mcK$, which is violated by the surface termination in the $x$-direction.}

\subsection{Surface signatures}\label{sec:spectra-surface}
For a photonic crystal, apart from the spectroscopic signatures 
of the bulk, the surface signatures can also be probed \cite{Yang:2018,Guo:2019}. In principle, one can obtain the energy spectra of the surface states by studying the photon scattering on a surface termination of the crystal. Recent research shows that for a general non-Hermitian Hamiltonian one must consider a non-Bloch bulk-boundary correspondence~\cite{Yao:2018,Yao:2018a,Yokomizo:2019}. Importantly, in cases where a non-Hermitian skin effect occurs~\cite{Lee:2016,Song:2019,Song:2019a}, it is crucial to define a Brillouin zone and the corresponding bulk states from an open boundary calculation~\cite{Yao:2018,Yao:2018a,Yokomizo:2019}. In our case, however, the model in Eq.~(\ref{eqn:lattice-model}) is set up to respect the regular bulk-boundary correspondence familiar from Hermitian models \TB{for open boundaries in the $x$ resp.~the $y$ direction}. 
%we can study the bulk-boundary correspondence in a more convenient way by using the special property of model Hamiltonian Eq.~\eqref{lattice-model}.
This is because our lattice Hamiltonian is obtained as a Hermitian matrix $H_{0}(\bs{k};m)$ multiplied by a phase factor $\e{\imi k_z/2}$. As a consequence, the Hermitian and the skew-Hermitian component of \TB{the lattice version of} $\mcH(\bs{k};m)$ %trivially
commute \TB{for open boundaries in the $x$ and in the $y$ directions}, implying the absence of the non-Hermitian skin effect \TB{on these boundaries}~\cite{Bergholtz:2019b}.

We therefore simplify the computation of the surface states for the model in Eq.~(\ref{eqn:lattice-model}) as follows. For a surface termination in the $x$ direction, the momenta along $y$ and $z$ directions are conserved, allowing us to define a surface Brillouin zone parametrized by $k_{y,z}\in[-\pi,\pi]$. However, instead of directly modelling the non-Hermitian Hamiltonian, we implement the Hermitian matrix $H_0(\bs{k};m)$. % for $k_z \in [-\pi,\pi]$. 
The surface spectra for the non-Hermitian model in Eq.~(\ref{eqn:lattice-model}) can then be obtained by multiplying the computed Hermitian spectra by a phase factor $\e{\imi k_z/2}$, which produces a simple half-twist in the complex plane as one traverses the surface Brillouin zone in the $k_z$ direction. 

Since the imprint of the phase factor on the surface density of states may be rather non-trivial, we visualize here the surface Fermi arcs by plotting the surface density of states of the Hermitian model $H_0(\bs{k};m)$ at zero energy. The result is shown for five values of parameter $m$ in Fig.~\ref{fig:conversion}(f--j). We remark that modelling the surface spectra is associated with some amount of freedom associated with boundary terms in the Hamiltonian. In our modelling, we add an extra term $H_\textrm{boundary} = \tfrac{1}{10}\sin\tfrac{k_z}{2}\sigma_z$ to the outermost layer of sites in the Hermitian model, in order to break the symmetry $C_{2y}\mcT: k_y \mapsto -k_y$ represented by $\sigma_x \mcK$. While the symmetry is respected by the bulk Hamiltonian in Eq.~(\ref{eqn:lattice-model}), it is explicitly broken by the open boundary in the $x$ direction. The inclusion of this term simplifies the form of the computed Fermi arcs, unpinning them from high-symmetry lines. Note that a surface band dispersing in the $k_z$ direction remains in the spectrum for $m\!>\!3$, cf.~Fig.~\ref{fig:conversion}(j). This band is topologically protected by the weak Chern parity~\cite{Wojcik:2019} in the $(k_x,k_y)$-plane, cf.~Sec.~\ref{sec:spectra-bulk}.
%number and the surface energy spectra is simply related to that of the Hermitian Hamiltonian by multiplying this factor. In this simpler setup, the  bulk-boundary correspondence is conventional. We plot the Fermi-arc states at zero energy corresponding to various values of $m$ in \figref{fig:conversion}(b). 

\section{Mathematical description}\label{sec:maths}
\subsection{Abe homotopy} \label{sec:Abe-general}
We describe topological charges of band nodes using homotopy groups~\cite{Bzdusek:2017,Fang:2015,Sun:2018}, %For simplicity, here we only consider two-band models, and we assume the absence of global symmetries (i.e.~symmetry class $\textrm{A}$ of Ref.~\cite{Kawabata:2018}). 
and we use $M$ to indicate the classifying space of Hamiltonians~\cite{Kitaev:2009}. The $p^\textrm{th}$ based homotopy group $\pi_p(M,\mathfrak{m})$ represents equivalence classes of continuous maps from a $p$-dimensional cube $I^p$ to $M$, such that the boundary $\partial I^p$ is mapped to a fixed base-point $\mathfrak{m}\!\in\!M$~\cite{Hatcher:2002}. The equivalence $f_1\!\sim\!f_2$ means that $f_1$ can be changed into $f_2$ by a continuous deformation that respects the boundary condition. When the boundary condition is relaxed such that $\partial I^p$ is mapped to a freely moving point $\mathfrak{m}$ in $M$, one speaks of a \emph{free} homotopy $\pi_p(M)$. Requiring that the whole boundary $\partial I^p$ is mapped onto a single point % $\mathfrak{m}$ 
%can be understood as identifying $\partial I^p$ with a single point, thus
effectively transforms $I^p$ into a $p$-sphere, $S^p$. Since band nodes of dimensions $d$ in a $D$-dimensional momentum space are naturally enclosed by spheres of dimension $D\!-\!d$, their topological charge is captured by homotopy group $\pi_{D-d}(M)$~\cite{Bzdusek:2017}.

\begin{figure}[b!]
    \centering
    \includegraphics[width=0.475\textwidth]{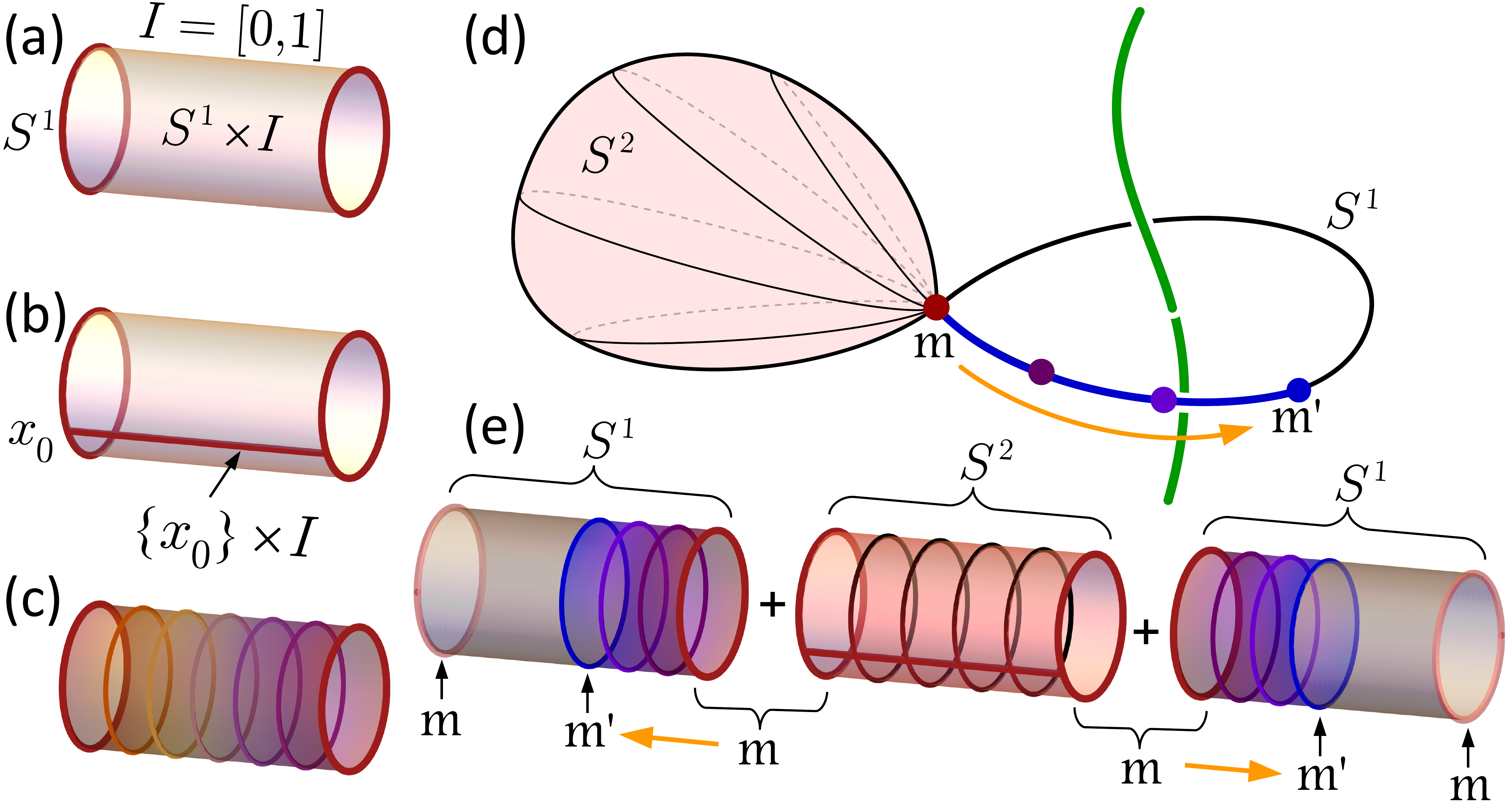}
    \caption{(a) Abe homotopy considers maps from a cylinder, $S^1 \!\times\! I$, with the boundary (red) mapped to the base-point $\mathfrak{m}$. (b) By further requiring a line segment $\{x_0\}\!\times\!I$ to be also mapped to the base-point, one recovers the second homotopy group, $\pi_2(M)$. (c) By narrowing attention to maps that only depend on the position along $I$, one recovers the first homotopy group, $\pi_1(M)$. (d) Abe homotopy allows us to compose elements of the second homotopy group (map on $S^2$) with elements of the first homotopy group (map on $S^1$). The green line indicates a band node. (e) We attach to the sphere $S^2$ based at $\mathfrak{m}$ a string from $\mathfrak{m}$ to $\mathfrak{m}'$. This corresponds to conjugating the cylinder representing the map on $S^2$ with a cylinder representing the map on the string. Attaching a \emph{closed} string (loop) based at $\mathfrak{m}$ corresponds to conjugating the element from $\pi_2(M)$ with an element from $\pi_1(M)$.}    
    \label{fig:Abe}
\end{figure}

The mathematical object that governs the observed non-commutative properties %of the topological charges 
of monopole charges near exceptional lines is known as the \emph{action of $\pi_{1}(M)$ on $\pi_{2}(M)$}~\cite{Tiwari:2019}. Returning back to the momentum-space picture in Fig.~\ref{fig:generics}, one can imagine continuously transforming the blue surface into two balloons, each containing only one Weyl point. The red surface is obtained by gluing the two balloons on the other side of the line defect (green). Carrying one of the balloons around the line induces an action of $\pi_1(M)$ (characterizing the closed path) on $\pi_2(M)$ (characterizing the transported balloon). %, but more complicated actions of $\pi_1$ on $\pi_2$ are also possible. 
The Alice string phenomenon corresponds to an action where odd elements of $\pi_1$ (corresponding e.g.~to a single line defect) reverses the sign of the elements in $\pi_2$. For example, in the case of uniaxial nematics~\cite{Volovik:1977}, the $\pi$-rotation of the order parameter on paths encircling a disclination line inverts the hedgehog configuration on the balloon~\cite{Alexander:2012}, thus manifesting the Alice nature of the disclination line~\cite{Volovik:1976}. The Alice string phenomenon can only arise if there is a difference between the free homotopy and the based homotopy~\cite{Arkowitz:2011}. %\TB{[TB: Casey, do we have a suitable math reference for this?]}

The action of $\pi_1$ on $\pi_2$ can be geometrically visualized and mathematically analyzed using the construction of Abe~\cite{Abe:1940}, who considered equivalence classes of maps from a cylinder $S^1 \!\times\! [0,1]$ to the target space $M$, such that the boundary $S^1\!\times\!\{0\} \,\cup\, S^1\!\times\!\{1\}$ is mapped to the base-point $\mathfrak{m}\!\in\! M$ [Fig.~\ref{fig:Abe}(a)]. By further requiring a segment $\{x\}\!\times\! [0,1]$ with a fixed $x\!\in\! S^1$ to be mapped to the base-point too, one reproduces the based homotopy group $\pi_2(M,\mathfrak{m})$ [Fig.~\ref{fig:Abe}(b)]. On the other hand, by limiting attention to ``stratified'' maps that only depend on the position $[0,1]$ \emph{along} the cylinder, one reproduces the based $\pi_1(M,\mathfrak{m})$ [Fig.~\ref{fig:Abe}(c)]. By stacking cylinders, we are able to combine elements of $\pi_1(M,\mathfrak{m})$ with elements of $\pi_2(M,\mathfrak{m})$ [Fig.~\ref{fig:Abe}(e)]. Especially, one can consider the effect of moving the base-point along a closed path in $M$, which corresponds to conjugating an element of $\pi_2(M,\mathfrak{m})$ with an element of $\pi_1(M,\mathfrak{m})$ [Fig.~\ref{fig:Abe}(d)]. The conjugation induces a map $\triangleright\!:\!\pi_1(M,\mathfrak{m})\!\to\!\textrm{Aut}[\pi_2(M,\mathfrak{m})]$, \emph{i.e.}~each element $g\!\in\!\pi_1(M,\mathfrak{m})$ is represented by an automorphism $\triangleright_g\!:\!\pi_2(M,\mathfrak{m})\!\to\!\pi_2(M,\mathfrak{m})$~\footnote{The compatibility further requires $\triangleright_g \!\circ\! \triangleright_h \!=\! \triangleright_{g\circ h}$. The collection $[\pi_1(M), \pi_2(M), \triangleright]$, together with an additional piece of data called the Postnikov class, form a mathematical structure called the \emph{fundamental 2-group of $M$}~\cite{Baez:2004,Ang:2018rls}.}, which is the sought action of $\pi_1$ on $\pi_2$. In the next section, we compute the action $\triangleright$ for non-Hermitian two-band models with no global symmetries (symmetry class $\textrm{A}$ of Ref.~\cite{Kawabata:2018}) to explain the Alice nature of exceptional lines.

\subsection{Topological charges revisited}\label{sec:topo-charges}

%\emph{Topological charges revisited.--}  
To explain % provide a mathematical underpinning of 
the non-commutative properties of topological charges observed in Sec.~\ref{sec:Chern-ambiguity},
%trivial braiding of band nodes, 
we identify the space $M$ of non-Hermitian two-band Hamiltonians, and we compute its based homotopy groups $\pi_{1}(M,\mathfrak{m})$, $\pi_{2}(M,\mathfrak{m})$ as well as the action $\triangleright$. Away from band nodes, the Hamiltonian exhibits two different eigenvalues. We therefore perform spectral flattening by making the Hamiltonian traceless (drop the term proportional to the identity matrix), and by normalizing the eigenvalues to absolute value $1$. This amounts to continuous deformations, % and without producing a band degeneracy, 
\emph{i.e.}~the procedure does not affect the band topology. To obtain homotopy groups, it is convenient~\footnote{Our follow-up work in Ref.~\cite{Wojcik:2019} adopts a different strategy, and expresses the same space as $M = S^2 \times S^1 /\ztwo$, where the quotient identifies antipodal points $(x,y)\sim (-x,-y)$ in $S^2\times S^1$.} to express $M$ as a coset space $\mathsf{G}/\mathsf{H}$ with $\mathsf{G}$ a simply connected group~\cite{Mermin:1979}. Then a mathematical theorem guarantees~\footnote{This follows from the long exact sequence of relative homotopy groups for pair $(\mathsf{G},\mathsf{H})$~\cite{Sun:2018,Hatcher:2002}.} that $\pi_{2}(M,\mathfrak{m})=\pi_{1}(\mathsf{H})$ and $\pi_{1}(M,\mathfrak{m})=\pi_{0}(\mathsf{H})$. 

To obtain the coset expression, we begin with the spectral decomposition of a generic two-band Hamiltonian $\mcH\in M$. Adopting the biorthogonal normalization of left and right eigenvectors~\cite{Brody:2013,Kunst:2018,Yao:2018a,Martinez:2018}, we obtain
\begin{equation}
\mcH = V^{-1} \cdot (\e{\imi \pi t} \sigma_z ) \cdot V,\label{eqn:decomp}
\end{equation}
where $\e{\imi\pi t}\sigma_z$ is a diagonal matrix containing the normalized eigenvalues, and $V$ is the matrix of the left eigenvectors of $\mcH$. Here, as a convention, we always rescale the eigenvectors such that $\det V \!=\! 1$, implying $V\!\in\!\mathsf{SL}(2,\cmplx)$. Therefore, the Hamiltonian can be encoded using two pieces of data, $(V,t)\!\in\! \mathsf{SL}(2,\cmplx)\!\times\! \reals \!\equiv\! \mathsf{G}$, which constitute a simply connected group~\cite{Procesi:2007} with composition rule $(V_1,t_1)\!\circ\! (V_2,t_2) \!=\! (V_1 \!\cdot\! V_2,t_1 \!+\! t_2)$. However, the decomposition of a non-Hermitian Hamiltonian into $(V,t)$ is not unique. On the one hand, the matrix $\mcH$ in Eq.~(\ref{eqn:decomp}) is invariant under rescaling the two eigenvectors separately by $(z,z^{-1})$ with $z\in\cmplx^\times$ (the complex plane without zero), as well as under shifting $t$ by an \emph{even} integer. This represents transformations
\begin{subequations}\label{eqn:stabilizer-action}
\begin{equation}
T_n(z):\; (V,t)\mapsto (R(z)\cdot V ,t+n) \qquad(\textrm{$n$ even})
\end{equation}
where $R(z) \!=\! \diag{(z,z^{-1})}$. On the other hand, we can flip the ordering of the eigenvectors if we appropriately reorder the eigenvalues $\e{\imi \pi t}\sigma_z$ by shifting $t$ by an \emph{odd} integer. This corresponds to transformations
\begin{equation}
T_n(z):\; (V,t)\mapsto (\imi \sigma_y\cdot R(z)\cdot V,t+n) \qquad(\textrm{$n$ odd}).
\end{equation}
\end{subequations}
Eqs.~(\ref{eqn:stabilizer-action}) represent left action on $\mathsf{G}$ by elements $T_n(z)$ [defined as $(R(z),n)$ for $n$ even, and as $(\imi\sigma_y\!\cdot\! R(z),n)$ for $n$ odd], which constitute a subgroup $\mathsf{H}\!<\!\mathsf{G}$. Therefore, the collection of all \emph{unique} non-Hermitian two-band Hamiltonians is the coset space $\mathsf{G}/\mathsf{H}$.

As a topological space, $\mathsf{H}$ is a disjoint union of many copies of $\cmplx^\times$ (one copy for each $n \!\in\! \intg$). It follows from Eqs.~(\ref{eqn:stabilizer-action}) that $T_{n_1}(z_1)\!\circ\! T_{n_2}(z_2) \!=\! T_{n_1 + n_2}(z)$ for some $z\!\in\!\cmplx^\times$, implying that connected components of $\mathsf{H}$ have a natural $\intg$-group structure. Therefore, $\pi_{1}(M)\!=\!\pi_{0}(\mathsf{H})\!=\!\intg$, which corresponds to the winding number. Furthermore, each disjoint component supports ``looping'' of $z$ around the origin of $\cmplx^\times$. Therefore, $\pi_{2}(M)\!=\!\pi_{1}(\mathsf{H})\!=\!\intg$, which corresponds to the Chern number. The details of the argument are discussed more carefully in Appendix~\ref{app:action}.

We are finally ready to compute the action of $\pi_1(M)$ on $\pi_2(M)$. According to Fig.~\ref{fig:Abe}(d), we should study the conjugation of elements in $\pi_2(M)$ [looping of the argument of $T_n(z)$] by elements in $\pi_1(M)$ [subscript of $T_n(z)$]. By considering all combinations of the parity of integers $n_1$ and $n_2$, we derive in Appendix~\ref{app:action} that
\begin{equation}
T_{n_1}(z_1)\circ T_{n_2}(z_2)\circ T_{n_1}(z_1)^{-1} = T_{n_2}\left(c\, z_2^{P(n_1)}\right),\label{eqn:action}
\end{equation}
where $P(n_1)\!=\!\pm 1$ is the parity of $n_1$, and $c\in\cmplx^\times$ is an unimportant factor that depends on $z_1$ and $n_{1,2}$. Since $z_2^{-1}$ loops around the origin of $\cmplx^\times$ opposite to $z_2$, we conclude that the Chern number flips sign if it is carried along a path with odd winding number. This is exactly what is anticipated for an Alice string. We emphasize that a path with a non-trivial winding number may exist even in the absence of exceptional lines, namely when the winding number along some direction of the BZ torus is odd, cf.~the model in Eq.~(\ref{eqn:lattice-model}). The result in Eq.~(\ref{eqn:action}) thus provides a mathematical underpinning of the Chern number transformations observed for all the models discussed in Sec.~\ref{sec:Chern-ambiguity} and~\ref{sec:spectra}.
%We present a lattice Hamiltonian with such a property in the Supplemental Material~\cite{Supp}.

\section{Conclusions and outlooks}\label{sec:conclude}
%\emph{Conclusions and outlooks.--} 
We have shown that band nodes with a Chern number braid non-trivially around exceptional lines in non-Hermitian systems, and that this interplay is naturally explained as an Alice string effect. While we have explicitly considered only two-band models, both the $\intg$-valued Chern number and the $\intg$-valued winding number are \emph{stable} topological invariants~\cite{Kawabata:2018}, therefore the non-trivial braiding of band nodes discussed here persists upon adding more bands. In fact, many-band models provide even richer possibilities. By traversing the Riemann-sheet band-structure near exceptional lines (which may now connect various pairs of bands), we can arbitrarily permute the ordering of the bands, and thus also of their Chern numbers. Especially, this allows us to move a Weyl point in-between a different pair of bands. This observation leads to a more general topological structure than the one considered in Ref.~\cite{Kawabata:2018}, which we develop in the work of Ref.~\cite{Wojcik:2019}. That work also considers the implications of the Alice string effect to the classification of topological band insulators.

We emphasize that a non-trivial action of $\pi_1(M)$ on $\pi_2(M)$ cannot arise for nodes in the stable limit of Hermitian systems. The observation from Ref.~\cite{Bzdusek:2017} is that if both of these homotopy groups are non-trivial, then $\pi_2(M)\!=\!\ztwo$, which does not support non-trivial automorphisms. There are only handful few-band Hermitian classes of Hamiltonians, very recently studied in Ref.~\cite{Tiwari:2019}, that enable a non-trivial braiding of monopole charges around line nodes~\cite{Fang:2015,Ahn:2018}. However, those examples are unstable against the inclusion of additional bands. Similarly, the non-Abelian reciprocal braiding of band nodes in Hermitian models discussed in recent Refs.~\cite{Wu:2018b,Ahn:2018b,Bouhon:2019} follows from a non-Abelian first homotopy group of the classifying space, rather than from a non-trivial action of $\pi_1$ and $\pi_2$, and its appearance also requires a subtle gap condition to arise. Therefore, the non-trivial braiding of band nodes in the stable limit %by the action of $\pi_1(M)$ on $\pi_2(M)$ 
constitutes a novel non-perturbative phenomenon enabled by non-Hermitian effects. 

Finally, we develop simple two-band lattice model to demonstrate the consequences of the topological structure. In particular, a pair of Weyl points in the band structure can be created and loop around the Brillouin zone torus by tuning one parameter. The pair cannot annihilate after this process. This will lead to bulk spectroscopic signature once the model is implemented in various possible platforms, such as acoustic meta-material or photonic crystals. 

%\TB{[TB: XiaoQi, anything to add to the conclusion? If yes, probably something about relation to experiments. The paragraphs above already comment on the generalization to the many band case, resp.~on the contrast with respect to Hermitian systems.]}

% Due to the high controllability and tunability, there have been experimental studies of both Weyl points and exceptional lines in photonic systems~\cite{Lu:2015,Chen:2016,Noh:2017,Cerjan:2018a}. In fact, the existing experimental techniques readily allow for inferring the topological charges of both exceptional nodes~\cite{Zhou:2018} and Weyl points~\cite{Wang:2017} in such systems. Because of the intrinsically non-Hermitian nature of photonic systems, we believe that a two-band model similar to Eq.~(\ref{model}) with one tuning parameter $\alpha$ (or to the lattice model presented in WHERE) could be realizable upon considering the gain and loss. Performing experiments analogous to those of Refs.~\cite{Lu:2015,Chen:2016,Noh:2017,Cerjan:2018a,Zhou:2018,Wang:2017} on such a model would provide experimental evidence for our predictions. Alternatively, one may consider implementing such a model in a system with synthetic dimension having gain and loss~\cite{Yuan:2018}. 

\begin{acknowledgments}
%\emph{Acknowledgments.--} 
We thank E.~Bergholtz, A.~Tiwari and M.~Xiao for helpful discussions, and R.-J.~Slager for providing comments on an early version of our draft. X.-Q.~S. was supported by the DOE Office of Science, Office of High Energy Physics, the grant DE-SC0019380, and in part by the Simons foundation. T.~B. was supported by the Gordon and Betty Moore Foundations EPiQS Initiative, Grant GBMF4302, and by the Ambizione Program of the Swiss National Science Foundation, Grant No.~185806. C.~C.~W. and S.~F. were supported by a Vannevar Bush Faculty Fellowship (Grant No. N00014-17-1-3030) from the U.S. Department of Defense.
\end{acknowledgments}

\appendix

\section{Derivation of Eq.~(\ref{eqn:action})}\label{app:action}

In this appendix, we complete the derivation of the action $\triangleright$ of $\pi_1$ on $\pi_2$, which we omitted in Sec.~\ref{sec:topo-charges}. Recall that $M$ is the space of $2\times 2$ Hamiltonians that are traceless and that have spectrum normalized to absolute value $1$. 
%We shall first review the formalism introduced by the main text. First, we express the space $M$ as a coset space $\mathsf{G}/\mathsf{H}$, where $\mathsf{G}$ is a simply connected Lie group and $\mathsf{H}$ is the stabilizer subgroup. Then, we use the coset expression and the computational algorithm described in Ref.~\cite{Mermin:1979} to derive $\pi_1(M)$ and $\pi_2(M)$, as well as the action $\triangleright$.
Following the decomposition Eq.~(\ref{eqn:decomp}), we identified any Hamiltonian in $M$ using $(V,t)\in \mathsf{SL}(2,\cmplx)\times \reals \equiv \mathsf{G}$. Furthermore, we argued that the stabilizer group $\mathsf{H}$ consists of the following elements in $\mathsf{G}$: $(R(z),n)$ for even $n$, and $(i\sigma_y\cdot R(z),n)$ for odd $n$, where $R(z)=\text{diag}(z,1/z)$ with $z$ being any complex number except of zero (which we indicate as $\cmplx/\{0\} \equiv \cmplx^\times$). We used a mathematical theorem from Ref.~\cite{Mermin:1979} to argue that $\pi_{2}(M,\mathfrak{m})=\pi_{1}(\mathsf{H})=\intg$ [i.e.~the Chern number on a 2-sphere corresponds to ``looping'' of the argument of $R(z)$ around the origin of $\cmplx^\times$], and that $\pi_{1}(M,\mathfrak{m})=\pi_{0}(\mathsf{H})=\intg$ [i.e.~the winding number on a 1-sphere corresponds to the connected component $n$ of the stabilizer $\mathsf{H}$]. 

The reduction of the information about the Hamiltonian from the coset space $\mathsf{G}/\mathsf{H}$ to the stabilizer (``gauge'') group $\mathsf{H}$ can be understood as follows. Given a circle or a $2$-sphere in $\bs{k}$-space, we represent it using the Abe cylinder as shown in Fig.~\ref{fig:Abe}(b) resp.~(c), with the corresponding points mapped to the base point $\mathfrak{m}$. The individual points of the cylinder are mapped into the space $\mathsf{G}$, with a non-uniqueness (i.e.~gauge) captured by the group $\mathsf{H}$. Since the exact image of the $p$-sphere (we assume $p\in\{1,2\}$) inside $\mathsf{G}$ is arbitrary up to the gauge transformations with $\mathsf{H}$, we need to study equivalence classes of maps to $\mathsf{G}/\mathsf{H}$, which are captured by the \emph{relative} homotopy group $\pi_p(\mathsf{G},\mathsf{H},\mathfrak{m})$~\cite{Hatcher:2002}. It follows from the simple connectedness of $\mathsf{G}$ and from a long exact sequence of homotopy groups~\cite{Mermin:1979} that the relative homotopy group is exactly equal to $\pi_{p-1}(\mathsf{H})$. This implies that the homotopy class of the Hamiltonian on the $p$-sphere is fully reproduced using an information inside the gauge group. More precisely, the topological information about the Hamiltonian on a $2$-sphere is encoded by a closed path in $\mathsf{H}$, and the topological information about the Hamiltonian on a circle is preserved by specifying a connected component of $\mathsf{H}$. Note also that a general element of the Abe homotopy [Fig.~\ref{fig:Abe}(a)] may produce paths in $\mathsf{H}$ with arbitrary value of $n_2$, while elements that correspond to based second homotopy group [Fig.~\ref{fig:Abe}(b)] correspond to paths with $n_2 \!=\! 0$ due to the constraint imposed at $\{x\}\!\times\![0,1]$.

According to Fig.~\ref{fig:Abe}(d), the action of $\pi_1$ on $\pi_2$ is found by studying the conjugation of elements in $\pi_{2}(M,\mathfrak{m})$ by elements in $\pi_{1}(M,\mathfrak{m})$. Following the reduction from $\mathsf{G}/\mathsf{H}$ to $\mathsf{H}$, the various equivalence classes of Hamiltonians in $\pi_{2}(M)$ are represented as topologically distinct loops inside $\mathsf{H}$, %i.e.~ as elements of $\pi_{1}(\mathsf{H})$. The looping inside $\mathsf{H}$ is captured by a 
i.e.~as a function that assigns to every point of $S^1$ some element $T_{n_2}(z_2)\in\mathsf{H}$ [namely some $(R(z_2),n_2)$ for even $n_2$ and some $(i\sigma_y\cdot R(z_2),n)$ for odd $n_2$.] 
% where $z_2$ is taken along a path that loops around the origin of the complex plane $\cmplx^\times$. 
%The $\pi_2$ invariant is the winding of $z_2$ around zero.
On the other hand, the elements in $\pi_{1}(M)$ are represented as points in the space of $\mathsf{H}$, i.e.~as some element $T_{n_1}(z_1)$. % with some $z_1\in\cmplx^\times$. 
Without loss of generality, we  set $z_1 = 1$ in our arguments below (we comment on the case of general $z_1$ at the very end). This identification allows us to explicitly compute $T_{n_1}(1)\circ T_{n_2}(z_2)\circ T_{n_1}(1)^{-1}$, which contains the information about the action $\triangleright$ of $\pi_1$ on $\pi_2$. 

Recall that the group $\mathsf{G}$ is a direct product of an Abelian group $\reals$ with addition as the group operation, and of a non-Abelian part $\mathsf{SL}(2,\cmplx)$. The Abelian part of $T_{n_2}(z_2)$ does not change upon conjugation by $T_{n_1}(1)$. Therefore, we only need to calculate the non-Abelian part of $T_{n_1}(1)\circ T_{n_2}(z_2)\circ T_{n_1}(1)^{-1}$. The calculation has to be split into several cases, corresponding to different parities of $n_1$ and $n_2$. First, for even $n_1$, $T_{n_1}(1)=(I_{2\times 2},n)$ commutes with $T_{n_2}(z_2)$, therefore
\begin{equation}
T_{n_1}(1)\circ T_{n_2}(z_2)\circ T_{n_1}(1)^{-1}=T_{n_2}(z_2) \quad\textrm{($n_1$ even)}, \label{eqn:action-n1-even}
\end{equation}
On the other hand, for odd $n_1$, we can use the commutation relations
\begin{equation}
\begin{split}
    &i\sigma_y\cdot R(z_2)=R(1/z_2)\cdot i\sigma_y\\
    &i\sigma_y\cdot (i\sigma_y\cdot R(z_2))=(i\sigma_y\cdot R(1/z_2))\cdot i\sigma_y
\end{split}
\end{equation}
to derive that for any $n_2$, we have
\begin{equation}
    T_{n_1}(1)\circ T_{n_2}(z_2)=T_{n_2}(1/z_2)\circ T_{n_1}(1)\quad\textrm{($n_1$ odd)}.
\end{equation}
It follows that
\begin{equation}
    T_{n_1}(1)\circ T_{n_2}(z_2)\circ T_{n_1}(1)^{-1}=T_{n_2}(1/z_2)\quad\textrm{($n_1$ odd)}.\label{eqn:action-n1-odd}
\end{equation}
The results in Eqs.~(\ref{eqn:action-n1-even}) and~(\ref{eqn:action-n1-odd}) can be compactly unified into a single equation
\begin{equation}
    T_{n_1}(1)\circ T_{n_2}(z_2)\circ T_{n_1}(1)^{-1}=T_{n_2}\left(z_2^{P(n_1)}\right),
\end{equation}
where $P(n_1)=\pm 1$ is the parity of $n_1$. Note that $1/z_2$ has opposite ``looping'' around the origin of $\cmplx^\times$ than $z_2$.
It follows that for odd $n_1$ [odd elements of $\pi_1(M,\mathfrak{m})$], the conjugation flips the sign of the $\pi_2(M)$ charge. 

For a general choice of $z_1$, one can explicitly compute for the four different combinations of parities of $n_1$ and $n_2$ the following results:
\begin{itemize}
\item If $n_{1}$ is even and $n_{2}$ is even, then
\begin{equation}
    T_{n_1}(z_1)\circ T_{n_2}(z_2)\circ T_{n_1}(z_1)^{-1}=T_{n_2}(z_2).
\end{equation}
\item If $n_{1}$ is even and $n_{2}$ is odd, then
\begin{equation}
    T_{n_1}(z_1)\circ T_{n_2}(z_2)\circ T_{n_1}(z_1)^{-1}=T_{n_2}(z_2/z_1^2).
\end{equation}
\item If $n_{1}$ is odd and $n_{2}$ is even, then
\begin{equation}
    T_{n_1}(z_1)\circ T_{n_2}(z_2)\circ T_{n_1}(z_1)^{-1}=T_{n_2}(1/z_2).
\end{equation}
\item If $n_{1}$ is odd and $n_{2}$ is odd, then
\begin{equation}
    T_{n_1}(z_1)\circ T_{n_2}(z_2)\circ T_{n_1}(z_1)^{-1}=T_{n_2}(z_1^2/z_2).
\end{equation}
\end{itemize}
The three equations are compactly summarized by Eq.~(\ref{eqn:action}) of Sec.~\ref{sec:topo-charges}. The derived action of $\pi_1$ on $\pi_2$ is compatible with a non-trivial Alice string phenomenon.

\section{\texorpdfstring{\\}{}Annihilating the Weyl points in Fig.~\ref{fig:model-2}(c)}\label{app:annihilation}

\subsection{Overview}

While the total Chern number of the two Weyl points in Fig.~\ref{fig:model-2}(c) is trivial, annihilating them at $\bs{k}=\bs{0}$ by appropriately amending the model in Eq.~(\ref{eqn:model-2}) is rather difficult. The underlying reason is the existence of an additional topological obstruction protected by the $C_2\mcT$ symmetry. This obstruction is subtle, because it corresponds to a relative-homotopy invariant on a hemisphere~\cite{Sun:2018}. In this appendix, we first analyze the relative-homotopy invariants relevant to the model in Eq.~(\ref{eqn:model-2}). Afterwards, we formulate a relaxed symmetry setting for which the Weyl points should be able to annihilate. Finally, we present an explicit continuous deformation of the Hamiltonian which annihilates the two Weyl points.

\subsection{The \texorpdfstring{$C_2\mcT$}{C2T} relative topology}

First, let us analyze the space of Hamiltonians inside the $C_2\mcT$-invariant plane $k_z = 0$. These are $2\times 2$ matrices that commute with the representation $\sigma_x \mcK$ of the symmetry. Dropping the term proportional to unit matrix, we find that $C_2\mcT$-symmetric Hamiltonians are 
\begin{equation}
\mcH(\bs{h}) = h_x \sigma_x + h_y\sigma_y - \imi h_z \sigma \label{eqn:C2T-Ham}
\end{equation}
with $h_{x,y,z}\in \reals$. We define the space $X$ to be \emph{non-degenerate} Hamiltonians of this form, and find that
\begin{equation}
X = \reals^3 \;\backslash \,\{\bs{h}\in\reals^3\;|\;h_x^2+h_y^2 = h_z^2\}.
\end{equation}
The condition on no degeneracy partitions the space $X$ into three disjoint components, namely 
\begin{eqnarray}
X^\pm &=& \{\bs{h}\in \reals^3 \;|\; \pm h_z > \sqrt{h_x^2+h_y^2}\}\qquad \\
\qquad\textrm{and}\quad X^0 &=& \{\bs{h}\in \reals^3 \;|\; \abs{h_z} < \sqrt{h_x^2+h_y^2}\}.\nonumber
\end{eqnarray}
The Hamiltonians inside the exceptional ring of Fig,~\ref{fig:model-2}(c) lie in the component $X^+$. Note that this component (the inside of a cone) is contractible to a point (i.e.~null-homotopic), therefore $\pi_n(X^+) = \triv$ for all $n\geq 1$.

We study topological obstruction on a hemisphere (i.e.~a deformed disc) with boundary lying \emph{inside} the exceptional ring in the $k_z\! =\! 0$ plane. This obstruction is captured by the second relative homotopy group $\pi_2(M,X^+)$~\cite{Sun:2018} where $M$ is the space of of \emph{all} traceless two-band Hamiltonians with non-degenerate spectrum, i.e.~the space considered in Sec.~\ref{sec:topo-charges}. It follows from the long exact sequence of homotopy groups~\cite{Hatcher:2002} and from the null-homotopy of $X^+$ that $\pi_2(M,X^+)\cong \pi_2(M)$ are isomorphic. Since $\pi_2(M) = \intg$ (the Chern number), it follows that \emph{one can define an integer topological invariant on the hemisphere}. We argue below that the hemisphere containing the Weyl point at $k_z = \sqrt{-\lambda}$ carries a non-trivial value of this invariant, which prevents the annihilation of the $C_2\mcT$-related Weyl points upon collision.

The relative homotopy invariant 
\begin{equation}
\pi_2(M,X^+)=\intg \label{eqn:C2T-invariant}
\end{equation}
on the hemisphere should be closely related to the Chern number. This may appear confusing at first, since the hemisphere is \emph{not} a closed surface. Nevertheless, the same finding was previously made by Ref.~\cite{Sun:2018} for two-band mirror-symmetric Hermitian Hamiltonians. In that case, a pair of mirror-ralated Weyl points (carrying opposite chirality) exhibit a non-trivial value of an analogous relative homotopy invariant, which was shown to induce a \emph{conversion} of a pair of mirror-related Weyl points into a nodal-line ring (rather then their annihilation). However, in the case of Ref.~\cite{Sun:2018} the Hamiltonian on the boundary of the hemisphere is essentially constant, which implies a quantized flow of Berry curvature (i.e. an integer Chern number) through the hemisphere. In contrast, this is \emph{not} true for the non-Hermitian Hamiltonian in Eq.~(\ref{eqn:C2T-Ham}) which has three free parameters. 

Nevertheless, the triviality $\pi_1(X^+)=\triv$ implies that the Hamiltonian on the boundary of the hemisphere for the non-Hermitian $C_2\mcT$-model can be continuously deformed into a constant without forming a band degeneracy along the way. Importantly, all such deformations to a constant are \emph{topologically equivalent}. For the specific case of Fig.~\ref{fig:model-2}(c), we can shrink the boundary of the hemisphere to a point in the middle of the exceptional ring. This transforms the hemisphere into a closed surface exhibiting a quantized Chern number, which is the integer invariant corresponding to Eq.~(\ref{eqn:C2T-invariant}). Importantly, this invariant is manifestly non-trivial for the situation in Fig.~\ref{fig:model-2}(c), because the resulting closed surface contains one Weyl point. Therefore, annihilation of the two Weyl points in the figure requires the breaking of the $C_2\mcT$ symmetry.

\subsection{Relaxing the symmetry setting}

The model in Eq.~(\ref{eqn:model-2}) exhibits more symmetry than just $C_2\mcT$. Especially, the model is symmetric under time-reversal $\mcT =  \sigma_y \mcK$, and under $\pi$-rotation around the $z$-axis $C_{2z} = - \imi \sigma_z$. In fact, the model respects the much stronger $\mathsf{SO}(2)$ rotation symmetry around the $z$-axis, which significantly simplifies the spectrum, and which we would therefore like to preserve. We therefore opt to remove $C_2\mcT$ symmetry by breaking $\mcT$.

The $\mathsf{SO}(2)$ symmetry makes it particularly elegant to define the Hamiltonian using operators in Eq.~(\ref{eqn:Lz-op}) which carry a well-defined angular momentum. Since the Hamiltonian should transform according to the trivial representation of $\mathsf{SO}(2)$, we only admit terms that have the same number of ``$+$'' and ``$-$'' constituents. For example, the first two terms in Eq.~(\ref{eqn:model-2}) are
\begin{equation}
k_x \sigma_x + k_y \sigma_y = k_+ \sigma_- + k_- \sigma_+.
\end{equation}
More generally, we admit for complex linear combinations of terms $(k_+ k_-)^a k_z^b \mathrm{s}$ where $a,b$ are non-negative integers and $\mathrm{s}\in\{k_+\sigma_-,k_-\sigma_+,\sigma_z\}$. The terms that break $\mcT$ correspond to combinations of an odd number of terms in the list
\begin{equation}
\{\imi, k_z, \sigma_z, (k_+\sigma_- - k_-\sigma_+), \imi(k_+\sigma_- + k_-\sigma_+) \}
\end{equation}
multiplied by real coefficients.

\subsection{Model for annihilating the Weyl points}

We amend the model in Eq.~(\ref{eqn:model-2}) as
\begin{eqnarray}
\mcH(\bs{k}) &=& [\alpha + (1-\alpha) k_+k_-](k_+\sigma_- + k_- \sigma_+)\nonumber \\
&\phantom{=}& + (k_z + \imi m)(k_z^2 - \beta k_+ k_- +\lambda)\sigma_z,\\
&\phantom{=}& + \gamma(\imi k_z -1)(k_+ \sigma_- - k_- \sigma_+) + \imi\delta \sigma_z \nonumber
\end{eqnarray}
where $\alpha,\beta,\gamma,\delta$ are four additional real coefficients. Note that only the term proportional to $\gamma$ breaks $C_2\mcT$. The situation in Fig.~\ref{fig:model-2}(c) corresponds to $\lambda = -1$, $m=+1$, $\alpha=+1$ and $\beta=\gamma=\delta=0$. 

We now deform the Hamiltonian in five steps:
\begin{enumerate}
\item Decrease $\alpha$ from $+1$ to $0$. This preserves the nodal structure plotted in Fig.~\ref{fig:model-2}(c).
\item Increase $\beta$ from $0$ to $+1$. This expands the radius of the exceptional ring at $k_z = 0$ from $1$ to $\approx 1.4656$, but otherwise it preserves the nodal structure.
\item Increase $\lambda$ from $-1$ to $0$. This has two effects. First, it brings the two Weyl points at $\bs{k}_\pm=(0,0,\pm 1)$ together at $\bs{k}=\bs{0}$. Second, it shrinks the radius of the exceptional ring at $k_z=0$ back to $1$.
\item We break $C_2\mcT$ by increasing $\gamma$ from $0$ to $+1$. This keeps the band touching at $\bs{k}=\bs{0}$, and it increases the radius of the exceptional ring from $1$ to $\approx 1.2720$.
\item Finally, we increase $\delta$ from $0$ to a small positive value. This opens a gap at $\bs{k}=\bs{0}$ (i.e.~it annihilates the two Weyl points), while the exceptional ring at $k_z = 0$ slightly shrinks. Setting $\delta$ specifically to $+1$ sets its radius to $1$.
\end{enumerate}
The only band degeneracy after the final step is the exceptional ring at $k_z = 0$. 

The exceptional ring can be shrunk to a single a Weyl point at $\bs{k} = \bs{0}$ by tuning the parameters as follows: 
\begin{itemize}
\item Increase $\lambda$ from $0$ to $+1$. 
\item Decrease $m$ from $+1$ to $0$.
\item Decrease $\beta$ from $+1$ to $0$.
\item Decrease $\gamma$ from $+1$ to $0$.
\item Increase $\alpha$ from $0$ to $+1$.
\item Decrease $\delta$ from $+1$ to $0$.
\end{itemize}
The final set of parameters match Fig.~\ref{fig:model-2}(a). We have thus gone a full cycle, converting the negative-chirality Weyl point of Fig.~\ref{fig:model-2}(b) through continuous deformation (and an intermediate non-Hermitian phase) into the positive-chirality Weyl point of Fig.~\ref{fig:model-2}(a).

\bibliography{bib}{}

%merlin.mbs apsrev4-1.bst 2010-07-25 4.21a (PWD, AO, DPC) hacked
%Control: key (0)
%Control: author (72) initials jnrlst
%Control: editor formatted (1) identically to author
%Control: production of article title (-1) disabled
%Control: page (0) single
%Control: year (1) truncated
%Control: production of eprint (0) enabled
\begin{thebibliography}{94}%
\makeatletter
\providecommand \@ifxundefined [1]{%
 \@ifx{#1\undefined}
}%
\providecommand \@ifnum [1]{%
 \ifnum #1\expandafter \@firstoftwo
 \else \expandafter \@secondoftwo
 \fi
}%
\providecommand \@ifx [1]{%
 \ifx #1\expandafter \@firstoftwo
 \else \expandafter \@secondoftwo
 \fi
}%
\providecommand \natexlab [1]{#1}%
\providecommand \enquote  [1]{``#1''}%
\providecommand \bibnamefont  [1]{#1}%
\providecommand \bibfnamefont [1]{#1}%
\providecommand \citenamefont [1]{#1}%
\providecommand \href@noop [0]{\@secondoftwo}%
\providecommand \href [0]{\begingroup \@sanitize@url \@href}%
\providecommand \@href[1]{\@@startlink{#1}\@@href}%
\providecommand \@@href[1]{\endgroup#1\@@endlink}%
\providecommand \@sanitize@url [0]{\catcode `\\12\catcode `\$12\catcode
  `\&12\catcode `\#12\catcode `\^12\catcode `\_12\catcode `\%12\relax}%
\providecommand \@@startlink[1]{}%
\providecommand \@@endlink[0]{}%
\providecommand \url  [0]{\begingroup\@sanitize@url \@url }%
\providecommand \@url [1]{\endgroup\@href {#1}{\urlprefix }}%
\providecommand \urlprefix  [0]{URL }%
\providecommand \Eprint [0]{\href }%
\providecommand \doibase [0]{http://dx.doi.org/}%
\providecommand \selectlanguage [0]{\@gobble}%
\providecommand \bibinfo  [0]{\@secondoftwo}%
\providecommand \bibfield  [0]{\@secondoftwo}%
\providecommand \translation [1]{[#1]}%
\providecommand \BibitemOpen [0]{}%
\providecommand \bibitemStop [0]{}%
\providecommand \bibitemNoStop [0]{.\EOS\space}%
\providecommand \EOS [0]{\spacefactor3000\relax}%
\providecommand \BibitemShut  [1]{\csname bibitem#1\endcsname}%
\let\auto@bib@innerbib\@empty
%</preamble>
\bibitem [{\citenamefont {Schwarz}(1982)}]{Schwarz:1982}%
  \BibitemOpen
  \bibfield  {author} {\bibinfo {author} {\bibfnamefont {A.~S.}\ \bibnamefont
  {Schwarz}},\ }\href {\doibase 10.1016/0550-3213(82)90190-0} {\bibfield
  {journal} {\bibinfo  {journal} {Nucl. Phys. B}\ }\textbf {\bibinfo {volume}
  {208}},\ \bibinfo {pages} {141} (\bibinfo {year} {1982})}\BibitemShut
  {NoStop}%
\bibitem [{\citenamefont {Volovik}(2003)}]{Volovik:2003}%
  \BibitemOpen
  \bibfield  {author} {\bibinfo {author} {\bibfnamefont {G.~E.}\ \bibnamefont
  {Volovik}},\ }\href@noop {} {\emph {\bibinfo {title} {The universe in a
  helium droplet}}},\ Vol.\ \bibinfo {volume} {117}\ (\bibinfo  {publisher}
  {Oxford University Press on Demand},\ \bibinfo {year} {2003})\BibitemShut
  {NoStop}%
\bibitem [{\citenamefont {Bucher}\ \emph {et~al.}(1992)\citenamefont {Bucher},
  \citenamefont {Lo},\ and\ \citenamefont {Preskill}}]{Bucher:1992}%
  \BibitemOpen
  \bibfield  {author} {\bibinfo {author} {\bibfnamefont {M.}~\bibnamefont
  {Bucher}}, \bibinfo {author} {\bibfnamefont {H.-K.}\ \bibnamefont {Lo}}, \
  and\ \bibinfo {author} {\bibfnamefont {J.}~\bibnamefont {Preskill}},\ }\href
  {\doibase https://doi.org/10.1016/0550-3213(92)90173-9} {\bibfield  {journal}
  {\bibinfo  {journal} {Nucl. Phys. B}\ }\textbf {\bibinfo {volume} {386}},\
  \bibinfo {pages} {3} (\bibinfo {year} {1992})}\BibitemShut {NoStop}%
\bibitem [{\citenamefont {Alford}\ \emph {et~al.}(1990)\citenamefont {Alford},
  \citenamefont {Benson}, \citenamefont {Coleman}, \citenamefont
  {March-Russell},\ and\ \citenamefont {Wilczek}}]{Alford:1990}%
  \BibitemOpen
  \bibfield  {author} {\bibinfo {author} {\bibfnamefont {M.~G.}\ \bibnamefont
  {Alford}}, \bibinfo {author} {\bibfnamefont {K.}~\bibnamefont {Benson}},
  \bibinfo {author} {\bibfnamefont {S.}~\bibnamefont {Coleman}}, \bibinfo
  {author} {\bibfnamefont {J.}~\bibnamefont {March-Russell}}, \ and\ \bibinfo
  {author} {\bibfnamefont {F.}~\bibnamefont {Wilczek}},\ }\href {\doibase
  10.1103/PhysRevLett.64.1632} {\bibfield  {journal} {\bibinfo  {journal}
  {Phys. Rev. Lett.}\ }\textbf {\bibinfo {volume} {64}},\ \bibinfo {pages}
  {1632} (\bibinfo {year} {1990})}\BibitemShut {NoStop}%
\bibitem [{\citenamefont {Leonhardt}\ and\ \citenamefont
  {Volovik}(2000)}]{Leonhardt:2000}%
  \BibitemOpen
  \bibfield  {author} {\bibinfo {author} {\bibfnamefont {U.}~\bibnamefont
  {Leonhardt}}\ and\ \bibinfo {author} {\bibfnamefont {G.~E.}\ \bibnamefont
  {Volovik}},\ }\href {\doibase 10.1134/1.1312008} {\bibfield  {journal}
  {\bibinfo  {journal} {JETP Lett.}\ }\textbf {\bibinfo {volume} {72}},\
  \bibinfo {pages} {46} (\bibinfo {year} {2000})}\BibitemShut {NoStop}%
\bibitem [{\citenamefont {Volovik}\ and\ \citenamefont
  {Mineev}(1976)}]{Volovik:1976}%
  \BibitemOpen
  \bibfield  {author} {\bibinfo {author} {\bibfnamefont {G.~E.}\ \bibnamefont
  {Volovik}}\ and\ \bibinfo {author} {\bibfnamefont {V.}~\bibnamefont
  {Mineev}},\ }\href {http://www.jetpletters.ac.ru/ps/1818/article_27785.shtml}
  {\bibfield  {journal} {\bibinfo  {journal} {JETP Lett.}\ }\textbf {\bibinfo
  {volume} {24}},\ \bibinfo {pages} {561} (\bibinfo {year} {1976})}\BibitemShut
  {NoStop}%
\bibitem [{\citenamefont {Benson}\ and\ \citenamefont
  {Imbo}(2004)}]{Benson:2004}%
  \BibitemOpen
  \bibfield  {author} {\bibinfo {author} {\bibfnamefont {K.~M.}\ \bibnamefont
  {Benson}}\ and\ \bibinfo {author} {\bibfnamefont {T.}~\bibnamefont {Imbo}},\
  }\href {\doibase 10.1103/PhysRevD.70.025005} {\bibfield  {journal} {\bibinfo
  {journal} {Phys. Rev. D}\ }\textbf {\bibinfo {volume} {70}},\ \bibinfo
  {pages} {025005} (\bibinfo {year} {2004})}\BibitemShut {NoStop}%
\bibitem [{\citenamefont {Mermin}(1979)}]{Mermin:1979}%
  \BibitemOpen
  \bibfield  {author} {\bibinfo {author} {\bibfnamefont {N.~D.}\ \bibnamefont
  {Mermin}},\ }\href {\doibase 10.1103/RevModPhys.51.591} {\bibfield  {journal}
  {\bibinfo  {journal} {Rev. Mod. Phys.}\ }\textbf {\bibinfo {volume} {51}},\
  \bibinfo {pages} {591} (\bibinfo {year} {1979})}\BibitemShut {NoStop}%
\bibitem [{\citenamefont {Bzdu\v{s}ek}\ and\ \citenamefont
  {Sigrist}(2017)}]{Bzdusek:2017}%
  \BibitemOpen
  \bibfield  {author} {\bibinfo {author} {\bibfnamefont {T.}~\bibnamefont
  {Bzdu\v{s}ek}}\ and\ \bibinfo {author} {\bibfnamefont {M.}~\bibnamefont
  {Sigrist}},\ }\href {\doibase 10.1103/PhysRevB.96.155105} {\bibfield
  {journal} {\bibinfo  {journal} {Phys. Rev. B}\ }\textbf {\bibinfo {volume}
  {96}},\ \bibinfo {pages} {155105} (\bibinfo {year} {2017})}\BibitemShut
  {NoStop}%
\bibitem [{\citenamefont {Fang}\ \emph {et~al.}(2015)\citenamefont {Fang},
  \citenamefont {Chen}, \citenamefont {Kee},\ and\ \citenamefont
  {Fu}}]{Fang:2015}%
  \BibitemOpen
  \bibfield  {author} {\bibinfo {author} {\bibfnamefont {C.}~\bibnamefont
  {Fang}}, \bibinfo {author} {\bibfnamefont {Y.}~\bibnamefont {Chen}}, \bibinfo
  {author} {\bibfnamefont {H.-Y.}\ \bibnamefont {Kee}}, \ and\ \bibinfo
  {author} {\bibfnamefont {L.~F.}\ \bibnamefont {Fu}},\ }\href {\doibase
  10.1103/PhysRevB.92.081201} {\bibfield  {journal} {\bibinfo  {journal} {Phys.
  Rev. B}\ }\textbf {\bibinfo {volume} {92}},\ \bibinfo {pages} {081201(R)}
  (\bibinfo {year} {2015})}\BibitemShut {NoStop}%
\bibitem [{\citenamefont {Sun}\ \emph {et~al.}(2018{\natexlab{a}})\citenamefont
  {Sun}, \citenamefont {Zhang},\ and\ \citenamefont {Bzdu\v{s}ek}}]{Sun:2018}%
  \BibitemOpen
  \bibfield  {author} {\bibinfo {author} {\bibfnamefont {X.-Q.}\ \bibnamefont
  {Sun}}, \bibinfo {author} {\bibfnamefont {S.-C.}\ \bibnamefont {Zhang}}, \
  and\ \bibinfo {author} {\bibfnamefont {T.}~\bibnamefont {Bzdu\v{s}ek}},\
  }\href {\doibase 10.1103/PhysRevLett.121.106402} {\bibfield  {journal}
  {\bibinfo  {journal} {Phys. Rev. Lett.}\ }\textbf {\bibinfo {volume} {121}},\
  \bibinfo {pages} {106402} (\bibinfo {year} {2018}{\natexlab{a}})}\BibitemShut
  {NoStop}%
\bibitem [{\citenamefont {Zhou}\ \emph {et~al.}(2018)\citenamefont {Zhou},
  \citenamefont {Peng}, \citenamefont {Yoon}, \citenamefont {Hsu},
  \citenamefont {Nelson}, \citenamefont {Fu}, \citenamefont {Joannopoulos},
  \citenamefont {Solja{\v{c}}i{\'c}},\ and\ \citenamefont {Zhen}}]{Zhou:2018}%
  \BibitemOpen
  \bibfield  {author} {\bibinfo {author} {\bibfnamefont {H.}~\bibnamefont
  {Zhou}}, \bibinfo {author} {\bibfnamefont {C.}~\bibnamefont {Peng}}, \bibinfo
  {author} {\bibfnamefont {Y.}~\bibnamefont {Yoon}}, \bibinfo {author}
  {\bibfnamefont {C.~W.}\ \bibnamefont {Hsu}}, \bibinfo {author} {\bibfnamefont
  {K.~A.}\ \bibnamefont {Nelson}}, \bibinfo {author} {\bibfnamefont
  {L.}~\bibnamefont {Fu}}, \bibinfo {author} {\bibfnamefont {J.~D.}\
  \bibnamefont {Joannopoulos}}, \bibinfo {author} {\bibfnamefont
  {M.}~\bibnamefont {Solja{\v{c}}i{\'c}}}, \ and\ \bibinfo {author}
  {\bibfnamefont {B.}~\bibnamefont {Zhen}},\ }\href {\doibase
  10.1126/science.aap9859} {\bibfield  {journal} {\bibinfo  {journal}
  {Science}\ }\textbf {\bibinfo {volume} {359}},\ \bibinfo {pages} {1009}
  (\bibinfo {year} {2018})}\BibitemShut {NoStop}%
\bibitem [{\citenamefont {Lu}\ \emph {et~al.}(2015)\citenamefont {Lu},
  \citenamefont {Wang}, \citenamefont {Ye}, \citenamefont {Ran}, \citenamefont
  {Fu}, \citenamefont {Joannopoulos},\ and\ \citenamefont {Solja{\v
  c}i{\'c}}}]{Lu:2015}%
  \BibitemOpen
  \bibfield  {author} {\bibinfo {author} {\bibfnamefont {L.}~\bibnamefont
  {Lu}}, \bibinfo {author} {\bibfnamefont {Z.}~\bibnamefont {Wang}}, \bibinfo
  {author} {\bibfnamefont {D.}~\bibnamefont {Ye}}, \bibinfo {author}
  {\bibfnamefont {L.}~\bibnamefont {Ran}}, \bibinfo {author} {\bibfnamefont
  {L.}~\bibnamefont {Fu}}, \bibinfo {author} {\bibfnamefont {J.~D.}\
  \bibnamefont {Joannopoulos}}, \ and\ \bibinfo {author} {\bibfnamefont
  {M.}~\bibnamefont {Solja{\v c}i{\'c}}},\ }\href {\doibase
  10.1126/science.aaa9273} {\bibfield  {journal} {\bibinfo  {journal}
  {Science}\ }\textbf {\bibinfo {volume} {349}},\ \bibinfo {pages} {622}
  (\bibinfo {year} {2015})}\BibitemShut {NoStop}%
\bibitem [{\citenamefont {Chen}\ \emph {et~al.}(2016)\citenamefont {Chen},
  \citenamefont {Xiao},\ and\ \citenamefont {Chan}}]{Chen:2016}%
  \BibitemOpen
  \bibfield  {author} {\bibinfo {author} {\bibfnamefont {W.-J.}\ \bibnamefont
  {Chen}}, \bibinfo {author} {\bibfnamefont {M.}~\bibnamefont {Xiao}}, \ and\
  \bibinfo {author} {\bibfnamefont {C.~T.}\ \bibnamefont {Chan}},\ }\href
  {https://doi.org/10.1038/ncomms13038} {\bibfield  {journal} {\bibinfo
  {journal} {Nat. Comm.}\ }\textbf {\bibinfo {volume} {7}},\ \bibinfo {pages}
  {13038} (\bibinfo {year} {2016})}\BibitemShut {NoStop}%
\bibitem [{\citenamefont {Noh}\ \emph {et~al.}(2017)\citenamefont {Noh},
  \citenamefont {Huang}, \citenamefont {Leykam}, \citenamefont {Chong},
  \citenamefont {Chen},\ and\ \citenamefont {Rechtsman}}]{Noh:2017}%
  \BibitemOpen
  \bibfield  {author} {\bibinfo {author} {\bibfnamefont {J.}~\bibnamefont
  {Noh}}, \bibinfo {author} {\bibfnamefont {S.}~\bibnamefont {Huang}}, \bibinfo
  {author} {\bibfnamefont {D.}~\bibnamefont {Leykam}}, \bibinfo {author}
  {\bibfnamefont {Y.~D.}\ \bibnamefont {Chong}}, \bibinfo {author}
  {\bibfnamefont {K.~P.}\ \bibnamefont {Chen}}, \ and\ \bibinfo {author}
  {\bibfnamefont {M.~C.}\ \bibnamefont {Rechtsman}},\ }\href
  {https://doi.org/10.1038/nphys4072} {\bibfield  {journal} {\bibinfo
  {journal} {Nat. Phys}\ }\textbf {\bibinfo {volume} {13}},\ \bibinfo {pages}
  {611} (\bibinfo {year} {2017})}\BibitemShut {NoStop}%
\bibitem [{\citenamefont {Cerjan}\ \emph
  {et~al.}(2018{\natexlab{a}})\citenamefont {Cerjan}, \citenamefont {Huang},
  \citenamefont {Chen}, \citenamefont {Chong},\ and\ \citenamefont
  {Rechtsman}}]{Cerjan:2018a}%
  \BibitemOpen
  \bibfield  {author} {\bibinfo {author} {\bibfnamefont {A.}~\bibnamefont
  {Cerjan}}, \bibinfo {author} {\bibfnamefont {S.}~\bibnamefont {Huang}},
  \bibinfo {author} {\bibfnamefont {K.~P.}\ \bibnamefont {Chen}}, \bibinfo
  {author} {\bibfnamefont {Y.}~\bibnamefont {Chong}}, \ and\ \bibinfo {author}
  {\bibfnamefont {M.~C.}\ \bibnamefont {Rechtsman}},\ }\href
  {https://arxiv.org/abs/1808.09541} {\bibfield  {journal} {\bibinfo  {journal}
  {arXiv:1808.09541}\ } (\bibinfo {year} {2018}{\natexlab{a}})}\BibitemShut
  {NoStop}%
\bibitem [{\citenamefont {Wang}\ \emph {et~al.}(2017)\citenamefont {Wang},
  \citenamefont {Xiao}, \citenamefont {Liu}, \citenamefont {Zhu},\ and\
  \citenamefont {Chan}}]{Wang:2017}%
  \BibitemOpen
  \bibfield  {author} {\bibinfo {author} {\bibfnamefont {Q.}~\bibnamefont
  {Wang}}, \bibinfo {author} {\bibfnamefont {M.}~\bibnamefont {Xiao}}, \bibinfo
  {author} {\bibfnamefont {H.}~\bibnamefont {Liu}}, \bibinfo {author}
  {\bibfnamefont {S.}~\bibnamefont {Zhu}}, \ and\ \bibinfo {author}
  {\bibfnamefont {C.~T.}\ \bibnamefont {Chan}},\ }\href {\doibase
  10.1103/PhysRevX.7.031032} {\bibfield  {journal} {\bibinfo  {journal} {Phys.
  Rev. X}\ }\textbf {\bibinfo {volume} {7}},\ \bibinfo {pages} {031032}
  (\bibinfo {year} {2017})}\BibitemShut {NoStop}%
\bibitem [{\citenamefont {Zhen}\ \emph {et~al.}(2015)\citenamefont {Zhen},
  \citenamefont {Hsu}, \citenamefont {Igarashi}, \citenamefont {Lu},
  \citenamefont {Kaminer}, \citenamefont {Pick}, \citenamefont {Chua},
  \citenamefont {Joannopoulos},\ and\ \citenamefont {Solja{\v
  c}i{\'c}}}]{Zhen:2015}%
  \BibitemOpen
  \bibfield  {author} {\bibinfo {author} {\bibfnamefont {B.}~\bibnamefont
  {Zhen}}, \bibinfo {author} {\bibfnamefont {C.~W.}\ \bibnamefont {Hsu}},
  \bibinfo {author} {\bibfnamefont {Y.}~\bibnamefont {Igarashi}}, \bibinfo
  {author} {\bibfnamefont {L.}~\bibnamefont {Lu}}, \bibinfo {author}
  {\bibfnamefont {I.}~\bibnamefont {Kaminer}}, \bibinfo {author} {\bibfnamefont
  {A.}~\bibnamefont {Pick}}, \bibinfo {author} {\bibfnamefont {S.-L.}\
  \bibnamefont {Chua}}, \bibinfo {author} {\bibfnamefont {J.~D.}\ \bibnamefont
  {Joannopoulos}}, \ and\ \bibinfo {author} {\bibfnamefont {M.}~\bibnamefont
  {Solja{\v c}i{\'c}}},\ }\href {https://doi.org/10.1038/nature14889}
  {\bibfield  {journal} {\bibinfo  {journal} {Nature}\ }\textbf {\bibinfo
  {volume} {525}},\ \bibinfo {pages} {354} (\bibinfo {year}
  {2015})}\BibitemShut {NoStop}%
\bibitem [{\citenamefont {Kozii}\ and\ \citenamefont {Fu}(2017)}]{Kozii:2017}%
  \BibitemOpen
  \bibfield  {author} {\bibinfo {author} {\bibfnamefont {V.}~\bibnamefont
  {Kozii}}\ and\ \bibinfo {author} {\bibfnamefont {L.}~\bibnamefont {Fu}},\
  }\href {https://arxiv.org/abs/1708.05841} {\bibfield  {journal} {\bibinfo
  {journal} {arXiv:1708.05841}\ } (\bibinfo {year} {2017})}\BibitemShut
  {NoStop}%
\bibitem [{\citenamefont {Shen}\ and\ \citenamefont {Fu}(2018)}]{Shen:2018}%
  \BibitemOpen
  \bibfield  {author} {\bibinfo {author} {\bibfnamefont {H.}~\bibnamefont
  {Shen}}\ and\ \bibinfo {author} {\bibfnamefont {L.}~\bibnamefont {Fu}},\
  }\href {\doibase 10.1103/PhysRevLett.121.026403} {\bibfield  {journal}
  {\bibinfo  {journal} {Phys. Rev. Lett.}\ }\textbf {\bibinfo {volume} {121}},\
  \bibinfo {pages} {026403} (\bibinfo {year} {2018})}\BibitemShut {NoStop}%
\bibitem [{\citenamefont {Wang}\ \emph {et~al.}(2019)\citenamefont {Wang},
  \citenamefont {Ruan},\ and\ \citenamefont {Zhang}}]{Wang:2019}%
  \BibitemOpen
  \bibfield  {author} {\bibinfo {author} {\bibfnamefont {H.}~\bibnamefont
  {Wang}}, \bibinfo {author} {\bibfnamefont {J.}~\bibnamefont {Ruan}}, \ and\
  \bibinfo {author} {\bibfnamefont {H.}~\bibnamefont {Zhang}},\ }\href
  {\doibase 10.1103/PhysRevB.99.075130} {\bibfield  {journal} {\bibinfo
  {journal} {Phys. Rev. B}\ }\textbf {\bibinfo {volume} {99}},\ \bibinfo
  {pages} {075130} (\bibinfo {year} {2019})}\BibitemShut {NoStop}%
\bibitem [{\citenamefont {Budich}\ \emph {et~al.}(2019)\citenamefont {Budich},
  \citenamefont {Carlstr\"om}, \citenamefont {Kunst},\ and\ \citenamefont
  {Bergholtz}}]{Budich:2019}%
  \BibitemOpen
  \bibfield  {author} {\bibinfo {author} {\bibfnamefont {J.~C.}\ \bibnamefont
  {Budich}}, \bibinfo {author} {\bibfnamefont {J.}~\bibnamefont {Carlstr\"om}},
  \bibinfo {author} {\bibfnamefont {F.~K.}\ \bibnamefont {Kunst}}, \ and\
  \bibinfo {author} {\bibfnamefont {E.~J.}\ \bibnamefont {Bergholtz}},\ }\href
  {\doibase 10.1103/PhysRevB.99.041406} {\bibfield  {journal} {\bibinfo
  {journal} {Phys. Rev. B}\ }\textbf {\bibinfo {volume} {99}},\ \bibinfo
  {pages} {041406} (\bibinfo {year} {2019})}\BibitemShut {NoStop}%
\bibitem [{\citenamefont {Yang}\ and\ \citenamefont {Hu}(2019)}]{Yang:2019}%
  \BibitemOpen
  \bibfield  {author} {\bibinfo {author} {\bibfnamefont {Z.}~\bibnamefont
  {Yang}}\ and\ \bibinfo {author} {\bibfnamefont {J.}~\bibnamefont {Hu}},\
  }\href {\doibase 10.1103/PhysRevB.99.081102} {\bibfield  {journal} {\bibinfo
  {journal} {Phys. Rev. B}\ }\textbf {\bibinfo {volume} {99}},\ \bibinfo
  {pages} {081102} (\bibinfo {year} {2019})}\BibitemShut {NoStop}%
\bibitem [{\citenamefont {Carlstr\"om}\ and\ \citenamefont
  {Bergholtz}(2018)}]{Carlstrom:2018}%
  \BibitemOpen
  \bibfield  {author} {\bibinfo {author} {\bibfnamefont {J.}~\bibnamefont
  {Carlstr\"om}}\ and\ \bibinfo {author} {\bibfnamefont {E.~J.}\ \bibnamefont
  {Bergholtz}},\ }\href {\doibase 10.1103/PhysRevA.98.042114} {\bibfield
  {journal} {\bibinfo  {journal} {Phys. Rev. A}\ }\textbf {\bibinfo {volume}
  {98}},\ \bibinfo {pages} {042114} (\bibinfo {year} {2018})}\BibitemShut
  {NoStop}%
\bibitem [{\citenamefont {Okugawa}\ and\ \citenamefont
  {Yokoyama}(2019)}]{Okugawa:2019}%
  \BibitemOpen
  \bibfield  {author} {\bibinfo {author} {\bibfnamefont {R.}~\bibnamefont
  {Okugawa}}\ and\ \bibinfo {author} {\bibfnamefont {T.}~\bibnamefont
  {Yokoyama}},\ }\href {\doibase 10.1103/PhysRevB.99.041202} {\bibfield
  {journal} {\bibinfo  {journal} {Phys. Rev. B}\ }\textbf {\bibinfo {volume}
  {99}},\ \bibinfo {pages} {041202} (\bibinfo {year} {2019})}\BibitemShut
  {NoStop}%
\bibitem [{\citenamefont {Moors}\ \emph {et~al.}(2019)\citenamefont {Moors},
  \citenamefont {Zyuzin}, \citenamefont {Zyuzin}, \citenamefont {Tiwari},\ and\
  \citenamefont {Schmidt}}]{Moors:2019}%
  \BibitemOpen
  \bibfield  {author} {\bibinfo {author} {\bibfnamefont {K.}~\bibnamefont
  {Moors}}, \bibinfo {author} {\bibfnamefont {A.~A.}\ \bibnamefont {Zyuzin}},
  \bibinfo {author} {\bibfnamefont {A.~Y.}\ \bibnamefont {Zyuzin}}, \bibinfo
  {author} {\bibfnamefont {R.~P.}\ \bibnamefont {Tiwari}}, \ and\ \bibinfo
  {author} {\bibfnamefont {T.~L.}\ \bibnamefont {Schmidt}},\ }\href {\doibase
  10.1103/PhysRevB.99.041116} {\bibfield  {journal} {\bibinfo  {journal} {Phys.
  Rev. B}\ }\textbf {\bibinfo {volume} {99}},\ \bibinfo {pages} {041116}
  (\bibinfo {year} {2019})}\BibitemShut {NoStop}%
\bibitem [{\citenamefont {Zyuzin}\ and\ \citenamefont
  {Zyuzin}(2018)}]{Zyuzin:2018}%
  \BibitemOpen
  \bibfield  {author} {\bibinfo {author} {\bibfnamefont {A.~A.}\ \bibnamefont
  {Zyuzin}}\ and\ \bibinfo {author} {\bibfnamefont {A.~Y.}\ \bibnamefont
  {Zyuzin}},\ }\href {\doibase 10.1103/PhysRevB.97.041203} {\bibfield
  {journal} {\bibinfo  {journal} {Phys. Rev. B}\ }\textbf {\bibinfo {volume}
  {97}},\ \bibinfo {pages} {041203} (\bibinfo {year} {2018})}\BibitemShut
  {NoStop}%
\bibitem [{\citenamefont {Yoshida}\ \emph {et~al.}(2019)\citenamefont
  {Yoshida}, \citenamefont {Peters}, \citenamefont {Kawakami},\ and\
  \citenamefont {Hatsugai}}]{Yoshida:2019}%
  \BibitemOpen
  \bibfield  {author} {\bibinfo {author} {\bibfnamefont {T.}~\bibnamefont
  {Yoshida}}, \bibinfo {author} {\bibfnamefont {R.}~\bibnamefont {Peters}},
  \bibinfo {author} {\bibfnamefont {N.}~\bibnamefont {Kawakami}}, \ and\
  \bibinfo {author} {\bibfnamefont {Y.}~\bibnamefont {Hatsugai}},\ }\href
  {\doibase 10.1103/PhysRevB.99.121101} {\bibfield  {journal} {\bibinfo
  {journal} {Phys. Rev. B}\ }\textbf {\bibinfo {volume} {99}},\ \bibinfo
  {pages} {121101} (\bibinfo {year} {2019})}\BibitemShut {NoStop}%
\bibitem [{\citenamefont {Zhou}\ and\ \citenamefont {Lee}(2019)}]{Zhou:2019}%
  \BibitemOpen
  \bibfield  {author} {\bibinfo {author} {\bibfnamefont {H.}~\bibnamefont
  {Zhou}}\ and\ \bibinfo {author} {\bibfnamefont {J.~Y.}\ \bibnamefont {Lee}},\
  }\href {\doibase 10.1103/PhysRevB.99.235112} {\bibfield  {journal} {\bibinfo
  {journal} {Phys. Rev. B}\ }\textbf {\bibinfo {volume} {99}},\ \bibinfo
  {pages} {235112} (\bibinfo {year} {2019})}\BibitemShut {NoStop}%
\bibitem [{\citenamefont {Zeuner}\ \emph {et~al.}(2015)\citenamefont {Zeuner},
  \citenamefont {Rechtsman}, \citenamefont {Plotnik}, \citenamefont {Lumer},
  \citenamefont {Nolte}, \citenamefont {Rudner}, \citenamefont {Segev},\ and\
  \citenamefont {Szameit}}]{Zeuner:2015}%
  \BibitemOpen
  \bibfield  {author} {\bibinfo {author} {\bibfnamefont {J.~M.}\ \bibnamefont
  {Zeuner}}, \bibinfo {author} {\bibfnamefont {M.~C.}\ \bibnamefont
  {Rechtsman}}, \bibinfo {author} {\bibfnamefont {Y.}~\bibnamefont {Plotnik}},
  \bibinfo {author} {\bibfnamefont {Y.}~\bibnamefont {Lumer}}, \bibinfo
  {author} {\bibfnamefont {S.}~\bibnamefont {Nolte}}, \bibinfo {author}
  {\bibfnamefont {M.~S.}\ \bibnamefont {Rudner}}, \bibinfo {author}
  {\bibfnamefont {M.}~\bibnamefont {Segev}}, \ and\ \bibinfo {author}
  {\bibfnamefont {A.}~\bibnamefont {Szameit}},\ }\href {\doibase
  10.1103/PhysRevLett.115.040402} {\bibfield  {journal} {\bibinfo  {journal}
  {Phys. Rev. Lett.}\ }\textbf {\bibinfo {volume} {115}},\ \bibinfo {pages}
  {040402} (\bibinfo {year} {2015})}\BibitemShut {NoStop}%
\bibitem [{\citenamefont {Xiao}\ \emph {et~al.}(2017)\citenamefont {Xiao},
  \citenamefont {Zhan}, \citenamefont {Bian}, \citenamefont {Wang},
  \citenamefont {Zhang}, \citenamefont {Wang}, \citenamefont {Li},
  \citenamefont {Mochizuki}, \citenamefont {Kim}, \citenamefont {Kawakami},
  \citenamefont {Yi}, \citenamefont {Obuse}, \citenamefont {Sanders},\ and\
  \citenamefont {Xue}}]{Xiao:2017}%
  \BibitemOpen
  \bibfield  {author} {\bibinfo {author} {\bibfnamefont {L.}~\bibnamefont
  {Xiao}}, \bibinfo {author} {\bibfnamefont {X.}~\bibnamefont {Zhan}}, \bibinfo
  {author} {\bibfnamefont {Z.~H.}\ \bibnamefont {Bian}}, \bibinfo {author}
  {\bibfnamefont {K.~K.}\ \bibnamefont {Wang}}, \bibinfo {author}
  {\bibfnamefont {X.}~\bibnamefont {Zhang}}, \bibinfo {author} {\bibfnamefont
  {X.~P.}\ \bibnamefont {Wang}}, \bibinfo {author} {\bibfnamefont
  {J.}~\bibnamefont {Li}}, \bibinfo {author} {\bibfnamefont {K.}~\bibnamefont
  {Mochizuki}}, \bibinfo {author} {\bibfnamefont {D.}~\bibnamefont {Kim}},
  \bibinfo {author} {\bibfnamefont {N.}~\bibnamefont {Kawakami}}, \bibinfo
  {author} {\bibfnamefont {W.}~\bibnamefont {Yi}}, \bibinfo {author}
  {\bibfnamefont {H.}~\bibnamefont {Obuse}}, \bibinfo {author} {\bibfnamefont
  {B.~C.}\ \bibnamefont {Sanders}}, \ and\ \bibinfo {author} {\bibfnamefont
  {P.}~\bibnamefont {Xue}},\ }\href {https://doi.org/10.1038/nphys4204}
  {\bibfield  {journal} {\bibinfo  {journal} {Nat. Phys.}\ }\textbf {\bibinfo
  {volume} {13}},\ \bibinfo {pages} {1117} (\bibinfo {year}
  {2017})}\BibitemShut {NoStop}%
\bibitem [{\citenamefont {Hu}\ \emph {et~al.}(2017)\citenamefont {Hu},
  \citenamefont {Wang}, \citenamefont {Shum},\ and\ \citenamefont
  {Chong}}]{Hu:2017}%
  \BibitemOpen
  \bibfield  {author} {\bibinfo {author} {\bibfnamefont {W.}~\bibnamefont
  {Hu}}, \bibinfo {author} {\bibfnamefont {H.}~\bibnamefont {Wang}}, \bibinfo
  {author} {\bibfnamefont {P.~P.}\ \bibnamefont {Shum}}, \ and\ \bibinfo
  {author} {\bibfnamefont {Y.~D.}\ \bibnamefont {Chong}},\ }\href {\doibase
  10.1103/PhysRevB.95.184306} {\bibfield  {journal} {\bibinfo  {journal} {Phys.
  Rev. B}\ }\textbf {\bibinfo {volume} {95}},\ \bibinfo {pages} {184306}
  (\bibinfo {year} {2017})}\BibitemShut {NoStop}%
\bibitem [{\citenamefont {Poli}\ \emph {et~al.}(2015)\citenamefont {Poli},
  \citenamefont {Bellec}, \citenamefont {Kuhl}, \citenamefont {Mortessagne},\
  and\ \citenamefont {Schomerus}}]{Poli:2015}%
  \BibitemOpen
  \bibfield  {author} {\bibinfo {author} {\bibfnamefont {C.}~\bibnamefont
  {Poli}}, \bibinfo {author} {\bibfnamefont {M.}~\bibnamefont {Bellec}},
  \bibinfo {author} {\bibfnamefont {U.}~\bibnamefont {Kuhl}}, \bibinfo {author}
  {\bibfnamefont {F.}~\bibnamefont {Mortessagne}}, \ and\ \bibinfo {author}
  {\bibfnamefont {H.}~\bibnamefont {Schomerus}},\ }\href
  {https://doi.org/10.1038/ncomms7710} {\bibfield  {journal} {\bibinfo
  {journal} {Nat. Comm.}\ }\textbf {\bibinfo {volume} {6}},\ \bibinfo {pages}
  {6710} (\bibinfo {year} {2015})}\BibitemShut {NoStop}%
\bibitem [{\citenamefont {Weimann}\ \emph {et~al.}(2016)\citenamefont
  {Weimann}, \citenamefont {Kremer}, \citenamefont {Plotnik}, \citenamefont
  {Lumer}, \citenamefont {Nolte}, \citenamefont {Makris}, \citenamefont
  {Segev}, \citenamefont {Rechtsman},\ and\ \citenamefont
  {Szameit}}]{Weimann:2016}%
  \BibitemOpen
  \bibfield  {author} {\bibinfo {author} {\bibfnamefont {S.}~\bibnamefont
  {Weimann}}, \bibinfo {author} {\bibfnamefont {M.}~\bibnamefont {Kremer}},
  \bibinfo {author} {\bibfnamefont {Y.}~\bibnamefont {Plotnik}}, \bibinfo
  {author} {\bibfnamefont {Y.}~\bibnamefont {Lumer}}, \bibinfo {author}
  {\bibfnamefont {S.}~\bibnamefont {Nolte}}, \bibinfo {author} {\bibfnamefont
  {K.~G.}\ \bibnamefont {Makris}}, \bibinfo {author} {\bibfnamefont
  {M.}~\bibnamefont {Segev}}, \bibinfo {author} {\bibfnamefont {M.~C.}\
  \bibnamefont {Rechtsman}}, \ and\ \bibinfo {author} {\bibfnamefont
  {A.}~\bibnamefont {Szameit}},\ }\href {https://doi.org/10.1038/nmat4811}
  {\bibfield  {journal} {\bibinfo  {journal} {Nat. Mater.}\ }\textbf {\bibinfo
  {volume} {16}},\ \bibinfo {pages} {433} (\bibinfo {year} {2016})}\BibitemShut
  {NoStop}%
\bibitem [{\citenamefont {Zhan}\ \emph {et~al.}(2017)\citenamefont {Zhan},
  \citenamefont {Xiao}, \citenamefont {Bian}, \citenamefont {Wang},
  \citenamefont {Qiu}, \citenamefont {Sanders}, \citenamefont {Yi},\ and\
  \citenamefont {Xue}}]{Zhan:2017}%
  \BibitemOpen
  \bibfield  {author} {\bibinfo {author} {\bibfnamefont {X.}~\bibnamefont
  {Zhan}}, \bibinfo {author} {\bibfnamefont {L.}~\bibnamefont {Xiao}}, \bibinfo
  {author} {\bibfnamefont {Z.}~\bibnamefont {Bian}}, \bibinfo {author}
  {\bibfnamefont {K.}~\bibnamefont {Wang}}, \bibinfo {author} {\bibfnamefont
  {X.}~\bibnamefont {Qiu}}, \bibinfo {author} {\bibfnamefont {B.~C.}\
  \bibnamefont {Sanders}}, \bibinfo {author} {\bibfnamefont {W.}~\bibnamefont
  {Yi}}, \ and\ \bibinfo {author} {\bibfnamefont {P.}~\bibnamefont {Xue}},\
  }\href {\doibase 10.1103/PhysRevLett.119.130501} {\bibfield  {journal}
  {\bibinfo  {journal} {Phys. Rev. Lett.}\ }\textbf {\bibinfo {volume} {119}},\
  \bibinfo {pages} {130501} (\bibinfo {year} {2017})}\BibitemShut {NoStop}%
\bibitem [{\citenamefont {Choi}\ \emph {et~al.}(2010)\citenamefont {Choi},
  \citenamefont {Kang}, \citenamefont {Lim}, \citenamefont {Kim}, \citenamefont
  {Kim}, \citenamefont {Lee},\ and\ \citenamefont {An}}]{Choi:2010}%
  \BibitemOpen
  \bibfield  {author} {\bibinfo {author} {\bibfnamefont {Y.}~\bibnamefont
  {Choi}}, \bibinfo {author} {\bibfnamefont {S.}~\bibnamefont {Kang}}, \bibinfo
  {author} {\bibfnamefont {S.}~\bibnamefont {Lim}}, \bibinfo {author}
  {\bibfnamefont {W.}~\bibnamefont {Kim}}, \bibinfo {author} {\bibfnamefont
  {J.-R.}\ \bibnamefont {Kim}}, \bibinfo {author} {\bibfnamefont {J.-H.}\
  \bibnamefont {Lee}}, \ and\ \bibinfo {author} {\bibfnamefont
  {K.}~\bibnamefont {An}},\ }\href {\doibase 10.1103/PhysRevLett.104.153601}
  {\bibfield  {journal} {\bibinfo  {journal} {Phys. Rev. Lett.}\ }\textbf
  {\bibinfo {volume} {104}},\ \bibinfo {pages} {153601} (\bibinfo {year}
  {2010})}\BibitemShut {NoStop}%
\bibitem [{\citenamefont {Papaj}\ \emph {et~al.}(2018)\citenamefont {Papaj},
  \citenamefont {Isobe},\ and\ \citenamefont {Fu}}]{Papaj:2018}%
  \BibitemOpen
  \bibfield  {author} {\bibinfo {author} {\bibfnamefont {M.}~\bibnamefont
  {Papaj}}, \bibinfo {author} {\bibfnamefont {H.}~\bibnamefont {Isobe}}, \ and\
  \bibinfo {author} {\bibfnamefont {L.}~\bibnamefont {Fu}},\ }\href
  {https://arxiv.org/abs/1802.00443} {\bibfield  {journal} {\bibinfo  {journal}
  {arXiv:1802.00443}\ } (\bibinfo {year} {2018})}\BibitemShut {NoStop}%
\bibitem [{\citenamefont {Zhong}\ \emph {et~al.}(2018)\citenamefont {Zhong},
  \citenamefont {Khajavikhan}, \citenamefont {Christodoulides},\ and\
  \citenamefont {El-Ganainy}}]{Zhong:2018}%
  \BibitemOpen
  \bibfield  {author} {\bibinfo {author} {\bibfnamefont {Q.}~\bibnamefont
  {Zhong}}, \bibinfo {author} {\bibfnamefont {M.}~\bibnamefont {Khajavikhan}},
  \bibinfo {author} {\bibfnamefont {D.~N.}\ \bibnamefont {Christodoulides}}, \
  and\ \bibinfo {author} {\bibfnamefont {R.}~\bibnamefont {El-Ganainy}},\
  }\href {\doibase 10.1038/s41467-018-07105-0} {\bibfield  {journal} {\bibinfo
  {journal} {Nature Communications}\ }\textbf {\bibinfo {volume} {9}},\
  \bibinfo {pages} {4808} (\bibinfo {year} {2018})}\BibitemShut {NoStop}%
\bibitem [{\citenamefont {Borgnia}\ \emph {et~al.}(2019)\citenamefont
  {Borgnia}, \citenamefont {Kruchkov},\ and\ \citenamefont
  {Slager}}]{Borgnia:2019}%
  \BibitemOpen
  \bibfield  {author} {\bibinfo {author} {\bibfnamefont {D.~S.}\ \bibnamefont
  {Borgnia}}, \bibinfo {author} {\bibfnamefont {A.~J.}\ \bibnamefont
  {Kruchkov}}, \ and\ \bibinfo {author} {\bibfnamefont {R.-J.}\ \bibnamefont
  {Slager}},\ }\href {https://arxiv.org/abs/1902.07217} {\bibfield  {journal}
  {\bibinfo  {journal} {arXiv:1902.07217}\ } (\bibinfo {year}
  {2019})}\BibitemShut {NoStop}%
\bibitem [{\citenamefont {Xu}\ \emph {et~al.}(2017)\citenamefont {Xu},
  \citenamefont {Wang},\ and\ \citenamefont {Duan}}]{Xu:2017}%
  \BibitemOpen
  \bibfield  {author} {\bibinfo {author} {\bibfnamefont {Y.}~\bibnamefont
  {Xu}}, \bibinfo {author} {\bibfnamefont {S.-T.}\ \bibnamefont {Wang}}, \ and\
  \bibinfo {author} {\bibfnamefont {L.-M.}\ \bibnamefont {Duan}},\ }\href
  {\doibase 10.1103/PhysRevLett.118.045701} {\bibfield  {journal} {\bibinfo
  {journal} {Phys. Rev. Lett.}\ }\textbf {\bibinfo {volume} {118}},\ \bibinfo
  {pages} {045701} (\bibinfo {year} {2017})}\BibitemShut {NoStop}%
\bibitem [{\citenamefont {Cerjan}\ \emph
  {et~al.}(2018{\natexlab{b}})\citenamefont {Cerjan}, \citenamefont {Xiao},
  \citenamefont {Yuan},\ and\ \citenamefont {Fan}}]{Cerjan:2018}%
  \BibitemOpen
  \bibfield  {author} {\bibinfo {author} {\bibfnamefont {A.}~\bibnamefont
  {Cerjan}}, \bibinfo {author} {\bibfnamefont {M.}~\bibnamefont {Xiao}},
  \bibinfo {author} {\bibfnamefont {L.}~\bibnamefont {Yuan}}, \ and\ \bibinfo
  {author} {\bibfnamefont {S.}~\bibnamefont {Fan}},\ }\href {\doibase
  10.1103/PhysRevB.97.075128} {\bibfield  {journal} {\bibinfo  {journal} {Phys.
  Rev. B}\ }\textbf {\bibinfo {volume} {97}},\ \bibinfo {pages} {075128}
  (\bibinfo {year} {2018}{\natexlab{b}})}\BibitemShut {NoStop}%
\bibitem [{\citenamefont {Carlstr\"om}\ \emph {et~al.}(2019)\citenamefont
  {Carlstr\"om}, \citenamefont {St\aa{}lhammar}, \citenamefont {Budich},\ and\
  \citenamefont {Bergholtz}}]{Carlstrom:2019}%
  \BibitemOpen
  \bibfield  {author} {\bibinfo {author} {\bibfnamefont {J.}~\bibnamefont
  {Carlstr\"om}}, \bibinfo {author} {\bibfnamefont {M.}~\bibnamefont
  {St\aa{}lhammar}}, \bibinfo {author} {\bibfnamefont {J.~C.}\ \bibnamefont
  {Budich}}, \ and\ \bibinfo {author} {\bibfnamefont {E.~J.}\ \bibnamefont
  {Bergholtz}},\ }\href {\doibase 10.1103/PhysRevB.99.161115} {\bibfield
  {journal} {\bibinfo  {journal} {Phys. Rev. B}\ }\textbf {\bibinfo {volume}
  {99}},\ \bibinfo {pages} {161115} (\bibinfo {year} {2019})}\BibitemShut
  {NoStop}%
\bibitem [{\citenamefont {Bergholtz}\ and\ \citenamefont
  {Budich}(2019)}]{Bergholtz:2019}%
  \BibitemOpen
  \bibfield  {author} {\bibinfo {author} {\bibfnamefont {E.~J.}\ \bibnamefont
  {Bergholtz}}\ and\ \bibinfo {author} {\bibfnamefont {J.~C.}\ \bibnamefont
  {Budich}},\ }\href {https://arxiv.org/abs/1903.12187} {\bibfield  {journal}
  {\bibinfo  {journal} {arXiv:1903.12187}\ } (\bibinfo {year}
  {2019})}\BibitemShut {NoStop}%
\bibitem [{\citenamefont {McClarty}\ and\ \citenamefont
  {Rau}(2019)}]{McClarty:2019}%
  \BibitemOpen
  \bibfield  {author} {\bibinfo {author} {\bibfnamefont {P.~A.}\ \bibnamefont
  {McClarty}}\ and\ \bibinfo {author} {\bibfnamefont {J.~G.}\ \bibnamefont
  {Rau}},\ }\href {\doibase 10.1103/PhysRevB.100.100405} {\bibfield  {journal}
  {\bibinfo  {journal} {Phys. Rev. B}\ }\textbf {\bibinfo {volume} {100}},\
  \bibinfo {pages} {100405} (\bibinfo {year} {2019})}\BibitemShut {NoStop}%
\bibitem [{\citenamefont {Kawabata}\ \emph {et~al.}(2019)\citenamefont
  {Kawabata}, \citenamefont {Bessho},\ and\ \citenamefont
  {Sato}}]{Kawabata:2019}%
  \BibitemOpen
  \bibfield  {author} {\bibinfo {author} {\bibfnamefont {K.}~\bibnamefont
  {Kawabata}}, \bibinfo {author} {\bibfnamefont {T.}~\bibnamefont {Bessho}}, \
  and\ \bibinfo {author} {\bibfnamefont {M.}~\bibnamefont {Sato}},\ }\href
  {https://arxiv.org/abs/1902.08479} {\bibfield  {journal} {\bibinfo  {journal}
  {arXiv:1902.08479}\ } (\bibinfo {year} {2019})}\BibitemShut {NoStop}%
\bibitem [{\citenamefont {Lee}\ \emph {et~al.}(2018)\citenamefont {Lee},
  \citenamefont {Li}, \citenamefont {Liu}, \citenamefont {Tai}, \citenamefont
  {Thomale},\ and\ \citenamefont {Zhang}}]{Lee:2018}%
  \BibitemOpen
  \bibfield  {author} {\bibinfo {author} {\bibfnamefont {C.~H.}\ \bibnamefont
  {Lee}}, \bibinfo {author} {\bibfnamefont {G.}~\bibnamefont {Li}}, \bibinfo
  {author} {\bibfnamefont {Y.}~\bibnamefont {Liu}}, \bibinfo {author}
  {\bibfnamefont {T.}~\bibnamefont {Tai}}, \bibinfo {author} {\bibfnamefont
  {R.}~\bibnamefont {Thomale}}, \ and\ \bibinfo {author} {\bibfnamefont
  {X.}~\bibnamefont {Zhang}},\ }\href@noop {} {\bibfield  {journal} {\bibinfo
  {journal} {arXiv preprint arXiv:1812.02011}\ } (\bibinfo {year}
  {2018})}\BibitemShut {NoStop}%
\bibitem [{\citenamefont {Lee}\ and\ \citenamefont {Thomale}(2019)}]{Lee:2019}%
  \BibitemOpen
  \bibfield  {author} {\bibinfo {author} {\bibfnamefont {C.~H.}\ \bibnamefont
  {Lee}}\ and\ \bibinfo {author} {\bibfnamefont {R.}~\bibnamefont {Thomale}},\
  }\href {\doibase 10.1103/PhysRevB.99.201103} {\bibfield  {journal} {\bibinfo
  {journal} {Phys. Rev. B}\ }\textbf {\bibinfo {volume} {99}},\ \bibinfo
  {pages} {201103} (\bibinfo {year} {2019})}\BibitemShut {NoStop}%
\bibitem [{\citenamefont {Bergholtz}\ \emph {et~al.}(2019)\citenamefont
  {Bergholtz}, \citenamefont {Budich},\ and\ \citenamefont
  {Kunst}}]{Bergholtz:2019b}%
  \BibitemOpen
  \bibfield  {author} {\bibinfo {author} {\bibfnamefont {E.~J.}\ \bibnamefont
  {Bergholtz}}, \bibinfo {author} {\bibfnamefont {J.~C.}\ \bibnamefont
  {Budich}}, \ and\ \bibinfo {author} {\bibfnamefont {F.~K.}\ \bibnamefont
  {Kunst}},\ }\href {https://arxiv.org/abs/1912.10048} {\bibfield  {journal}
  {\bibinfo  {journal} {arXiv:1912.10048}\ } (\bibinfo {year}
  {2019})}\BibitemShut {NoStop}%
\bibitem [{\citenamefont {Yoshida}\ \emph {et~al.}(2020)\citenamefont
  {Yoshida}, \citenamefont {Peters}, \citenamefont {Kawakami},\ and\
  \citenamefont {Hatsugai}}]{Yoshida:2020}%
  \BibitemOpen
  \bibfield  {author} {\bibinfo {author} {\bibfnamefont {T.}~\bibnamefont
  {Yoshida}}, \bibinfo {author} {\bibfnamefont {R.}~\bibnamefont {Peters}},
  \bibinfo {author} {\bibfnamefont {N.}~\bibnamefont {Kawakami}}, \ and\
  \bibinfo {author} {\bibfnamefont {Y.}~\bibnamefont {Hatsugai}},\ }\href
  {https://arxiv.org/abs/2002.11265} {\bibfield  {journal} {\bibinfo  {journal}
  {arXiv:2002.11265}\ } (\bibinfo {year} {2020})}\BibitemShut {NoStop}%
\bibitem [{\citenamefont {Murakami}(2007)}]{Murakami:2007}%
  \BibitemOpen
  \bibfield  {author} {\bibinfo {author} {\bibfnamefont {S.}~\bibnamefont
  {Murakami}},\ }\href {\doibase 10.1088/1367-2630/9/9/356} {\bibfield
  {journal} {\bibinfo  {journal} {New J. Phys.}\ }\textbf {\bibinfo {volume}
  {9}},\ \bibinfo {pages} {356} (\bibinfo {year} {2007})}\BibitemShut {NoStop}%
\bibitem [{\citenamefont {Wan}\ \emph {et~al.}(2011)\citenamefont {Wan},
  \citenamefont {Turner}, \citenamefont {Vishwanath},\ and\ \citenamefont
  {Savrasov}}]{Wan:2011}%
  \BibitemOpen
  \bibfield  {author} {\bibinfo {author} {\bibfnamefont {X.}~\bibnamefont
  {Wan}}, \bibinfo {author} {\bibfnamefont {A.~M.}\ \bibnamefont {Turner}},
  \bibinfo {author} {\bibfnamefont {A.}~\bibnamefont {Vishwanath}}, \ and\
  \bibinfo {author} {\bibfnamefont {S.~Y.}\ \bibnamefont {Savrasov}},\ }\href
  {\doibase 10.1103/PhysRevB.83.205101} {\bibfield  {journal} {\bibinfo
  {journal} {Phys. Rev. B}\ }\textbf {\bibinfo {volume} {83}},\ \bibinfo
  {pages} {205101} (\bibinfo {year} {2011})}\BibitemShut {NoStop}%
\bibitem [{\citenamefont {Shen}\ \emph {et~al.}(2018)\citenamefont {Shen},
  \citenamefont {Zhen},\ and\ \citenamefont {Fu}}]{Shen:2018a}%
  \BibitemOpen
  \bibfield  {author} {\bibinfo {author} {\bibfnamefont {H.}~\bibnamefont
  {Shen}}, \bibinfo {author} {\bibfnamefont {B.}~\bibnamefont {Zhen}}, \ and\
  \bibinfo {author} {\bibfnamefont {L.}~\bibnamefont {Fu}},\ }\href {\doibase
  10.1103/PhysRevLett.120.146402} {\bibfield  {journal} {\bibinfo  {journal}
  {Phys. Rev. Lett.}\ }\textbf {\bibinfo {volume} {120}},\ \bibinfo {pages}
  {146402} (\bibinfo {year} {2018})}\BibitemShut {NoStop}%
\bibitem [{\citenamefont {Yao}\ \emph {et~al.}(2018)\citenamefont {Yao},
  \citenamefont {Song},\ and\ \citenamefont {Wang}}]{Yao:2018}%
  \BibitemOpen
  \bibfield  {author} {\bibinfo {author} {\bibfnamefont {S.}~\bibnamefont
  {Yao}}, \bibinfo {author} {\bibfnamefont {F.}~\bibnamefont {Song}}, \ and\
  \bibinfo {author} {\bibfnamefont {Z.}~\bibnamefont {Wang}},\ }\href {\doibase
  10.1103/PhysRevLett.121.136802} {\bibfield  {journal} {\bibinfo  {journal}
  {Phys. Rev. Lett.}\ }\textbf {\bibinfo {volume} {121}},\ \bibinfo {pages}
  {136802} (\bibinfo {year} {2018})}\BibitemShut {NoStop}%
\bibitem [{\citenamefont {Gong}\ \emph {et~al.}(2018)\citenamefont {Gong},
  \citenamefont {Ashida}, \citenamefont {Kawabata}, \citenamefont {Takasan},
  \citenamefont {Higashikawa},\ and\ \citenamefont {Ueda}}]{Gong:2018}%
  \BibitemOpen
  \bibfield  {author} {\bibinfo {author} {\bibfnamefont {Z.}~\bibnamefont
  {Gong}}, \bibinfo {author} {\bibfnamefont {Y.}~\bibnamefont {Ashida}},
  \bibinfo {author} {\bibfnamefont {K.}~\bibnamefont {Kawabata}}, \bibinfo
  {author} {\bibfnamefont {K.}~\bibnamefont {Takasan}}, \bibinfo {author}
  {\bibfnamefont {S.}~\bibnamefont {Higashikawa}}, \ and\ \bibinfo {author}
  {\bibfnamefont {M.}~\bibnamefont {Ueda}},\ }\href {\doibase
  10.1103/PhysRevX.8.031079} {\bibfield  {journal} {\bibinfo  {journal} {Phys.
  Rev. X}\ }\textbf {\bibinfo {volume} {8}},\ \bibinfo {pages} {031079}
  (\bibinfo {year} {2018})}\BibitemShut {NoStop}%
\bibitem [{\citenamefont {Song}\ \emph
  {et~al.}(2019{\natexlab{a}})\citenamefont {Song}, \citenamefont {Yao},\ and\
  \citenamefont {Wang}}]{Song:2019}%
  \BibitemOpen
  \bibfield  {author} {\bibinfo {author} {\bibfnamefont {F.}~\bibnamefont
  {Song}}, \bibinfo {author} {\bibfnamefont {S.}~\bibnamefont {Yao}}, \ and\
  \bibinfo {author} {\bibfnamefont {Z.}~\bibnamefont {Wang}},\ }\href {\doibase
  10.1103/PhysRevLett.123.246801} {\bibfield  {journal} {\bibinfo  {journal}
  {Phys. Rev. Lett.}\ }\textbf {\bibinfo {volume} {123}},\ \bibinfo {pages}
  {246801} (\bibinfo {year} {2019}{\natexlab{a}})}\BibitemShut {NoStop}%
\bibitem [{\citenamefont {Ghatak}\ and\ \citenamefont
  {Das}(2019)}]{Ghatak:2019}%
  \BibitemOpen
  \bibfield  {author} {\bibinfo {author} {\bibfnamefont {A.}~\bibnamefont
  {Ghatak}}\ and\ \bibinfo {author} {\bibfnamefont {T.}~\bibnamefont {Das}},\
  }\href {\doibase 10.1088/1361-648X/ab11b3} {\bibfield  {journal} {\bibinfo
  {journal} {J. Phys. Condens. Matter}\ }\textbf {\bibinfo {volume} {31}},\
  \bibinfo {pages} {263001} (\bibinfo {year} {2019})}\BibitemShut {NoStop}%
\bibitem [{\citenamefont {Kawabata}\ \emph {et~al.}(2018)\citenamefont
  {Kawabata}, \citenamefont {Shiozaki}, \citenamefont {Ueda},\ and\
  \citenamefont {Sato}}]{Kawabata:2018}%
  \BibitemOpen
  \bibfield  {author} {\bibinfo {author} {\bibfnamefont {K.}~\bibnamefont
  {Kawabata}}, \bibinfo {author} {\bibfnamefont {K.}~\bibnamefont {Shiozaki}},
  \bibinfo {author} {\bibfnamefont {M.}~\bibnamefont {Ueda}}, \ and\ \bibinfo
  {author} {\bibfnamefont {M.}~\bibnamefont {Sato}},\ }\href
  {https://arxiv.org/abs/1812.09133} {\bibfield  {journal} {\bibinfo  {journal}
  {arXiv:1812.09133}\ } (\bibinfo {year} {2018})}\BibitemShut {NoStop}%
\bibitem [{\citenamefont {Sun}\ \emph {et~al.}(2018{\natexlab{b}})\citenamefont
  {Sun}, \citenamefont {Xiao}, \citenamefont {Bzdu\v{s}ek}, \citenamefont
  {Zhang},\ and\ \citenamefont {Fan}}]{Sun:2018b}%
  \BibitemOpen
  \bibfield  {author} {\bibinfo {author} {\bibfnamefont {X.-Q.}\ \bibnamefont
  {Sun}}, \bibinfo {author} {\bibfnamefont {M.}~\bibnamefont {Xiao}}, \bibinfo
  {author} {\bibfnamefont {T.}~\bibnamefont {Bzdu\v{s}ek}}, \bibinfo {author}
  {\bibfnamefont {S.-C.}\ \bibnamefont {Zhang}}, \ and\ \bibinfo {author}
  {\bibfnamefont {S.}~\bibnamefont {Fan}},\ }\href {\doibase
  10.1103/PhysRevLett.121.196401} {\bibfield  {journal} {\bibinfo  {journal}
  {Phys. Rev. Lett.}\ }\textbf {\bibinfo {volume} {121}},\ \bibinfo {pages}
  {196401} (\bibinfo {year} {2018}{\natexlab{b}})}\BibitemShut {NoStop}%
\bibitem [{\citenamefont {Xu}\ \emph {et~al.}(2016)\citenamefont {Xu},
  \citenamefont {Mason}, \citenamefont {Jiang},\ and\ \citenamefont
  {Harris}}]{Xu:2016}%
  \BibitemOpen
  \bibfield  {author} {\bibinfo {author} {\bibfnamefont {H.}~\bibnamefont
  {Xu}}, \bibinfo {author} {\bibfnamefont {D.}~\bibnamefont {Mason}}, \bibinfo
  {author} {\bibfnamefont {L.}~\bibnamefont {Jiang}}, \ and\ \bibinfo {author}
  {\bibfnamefont {J.~G. E.~H.}\ \bibnamefont {Harris}},\ }\href
  {https://doi.org/10.1038/nature18604} {\bibfield  {journal} {\bibinfo
  {journal} {Nature}\ }\textbf {\bibinfo {volume} {537}},\ \bibinfo {pages}
  {80} (\bibinfo {year} {2016})}\BibitemShut {NoStop}%
\bibitem [{\citenamefont {Doppler}\ \emph {et~al.}(2016)\citenamefont
  {Doppler}, \citenamefont {Mailybaev}, \citenamefont {B\"{o}hm}, \citenamefont
  {Kuhl}, \citenamefont {Girschik}, \citenamefont {Libisch}, \citenamefont
  {Milburn}, \citenamefont {Rabl}, \citenamefont {Moiseyev},\ and\
  \citenamefont {Rotter}}]{Doppler:2016}%
  \BibitemOpen
  \bibfield  {author} {\bibinfo {author} {\bibfnamefont {J.}~\bibnamefont
  {Doppler}}, \bibinfo {author} {\bibfnamefont {A.~A.}\ \bibnamefont
  {Mailybaev}}, \bibinfo {author} {\bibfnamefont {J.}~\bibnamefont {B\"{o}hm}},
  \bibinfo {author} {\bibfnamefont {U.}~\bibnamefont {Kuhl}}, \bibinfo {author}
  {\bibfnamefont {A.}~\bibnamefont {Girschik}}, \bibinfo {author}
  {\bibfnamefont {F.}~\bibnamefont {Libisch}}, \bibinfo {author} {\bibfnamefont
  {T.~J.}\ \bibnamefont {Milburn}}, \bibinfo {author} {\bibfnamefont
  {P.}~\bibnamefont {Rabl}}, \bibinfo {author} {\bibfnamefont {N.}~\bibnamefont
  {Moiseyev}}, \ and\ \bibinfo {author} {\bibfnamefont {S.}~\bibnamefont
  {Rotter}},\ }\href {https://doi.org/10.1038/nature18605} {\bibfield
  {journal} {\bibinfo  {journal} {Nature}\ }\textbf {\bibinfo {volume} {537}},\
  \bibinfo {pages} {76} (\bibinfo {year} {2016})}\BibitemShut {NoStop}%
\bibitem [{\citenamefont {Berry}(2004)}]{Berry:2004}%
  \BibitemOpen
  \bibfield  {author} {\bibinfo {author} {\bibfnamefont {M.~V.}\ \bibnamefont
  {Berry}},\ }\href {\doibase 10.1023/B:CJOP.0000044002.05657.04} {\bibfield
  {journal} {\bibinfo  {journal} {Czech. J. Phys.}\ }\textbf {\bibinfo {volume}
  {54}},\ \bibinfo {pages} {1039} (\bibinfo {year} {2004})}\BibitemShut
  {NoStop}%
\bibitem [{\citenamefont {Wojcik}\ \emph {et~al.}(2020)\citenamefont {Wojcik},
  \citenamefont {Sun}, \citenamefont {Bzdu\ifmmode~\check{s}\else
  \v{s}\fi{}ek},\ and\ \citenamefont {Fan}}]{Wojcik:2019}%
  \BibitemOpen
  \bibfield  {author} {\bibinfo {author} {\bibfnamefont {C.~C.}\ \bibnamefont
  {Wojcik}}, \bibinfo {author} {\bibfnamefont {X.-Q.}\ \bibnamefont {Sun}},
  \bibinfo {author} {\bibfnamefont {T.~c.~v.}\ \bibnamefont
  {Bzdu\ifmmode~\check{s}\else \v{s}\fi{}ek}}, \ and\ \bibinfo {author}
  {\bibfnamefont {S.}~\bibnamefont {Fan}},\ }\href {\doibase
  10.1103/PhysRevB.101.205417} {\bibfield  {journal} {\bibinfo  {journal}
  {Phys. Rev. B}\ }\textbf {\bibinfo {volume} {101}},\ \bibinfo {pages}
  {205417} (\bibinfo {year} {2020})}\BibitemShut {NoStop}%
\bibitem [{Note1()}]{Note1}%
  \BibitemOpen
  \bibinfo {note} {The definition of the Wilson loop operator and the
  corresponding Berry phase depends on the normalization of the states in
  non-Hermitian systems. The convention of this paper corresponds to the
  left-right Berry curvature in Ref.~\cite {Shen:2018a}}\BibitemShut {NoStop}%
\bibitem [{\citenamefont {Gresch}\ \emph {et~al.}(2017)\citenamefont {Gresch},
  \citenamefont {Aut\`es}, \citenamefont {Yazyev}, \citenamefont {Troyer},
  \citenamefont {Vanderbilt}, \citenamefont {Bernevig},\ and\ \citenamefont
  {Soluyanov}}]{Gresch:2017}%
  \BibitemOpen
  \bibfield  {author} {\bibinfo {author} {\bibfnamefont {D.}~\bibnamefont
  {Gresch}}, \bibinfo {author} {\bibfnamefont {G.}~\bibnamefont {Aut\`es}},
  \bibinfo {author} {\bibfnamefont {O.~V.}\ \bibnamefont {Yazyev}}, \bibinfo
  {author} {\bibfnamefont {M.}~\bibnamefont {Troyer}}, \bibinfo {author}
  {\bibfnamefont {D.}~\bibnamefont {Vanderbilt}}, \bibinfo {author}
  {\bibfnamefont {B.~A.}\ \bibnamefont {Bernevig}}, \ and\ \bibinfo {author}
  {\bibfnamefont {A.~A.}\ \bibnamefont {Soluyanov}},\ }\href {\doibase
  10.1103/PhysRevB.95.075146} {\bibfield  {journal} {\bibinfo  {journal} {Phys.
  Rev. B}\ }\textbf {\bibinfo {volume} {95}},\ \bibinfo {pages} {075146}
  (\bibinfo {year} {2017})}\BibitemShut {NoStop}%
\bibitem [{\citenamefont {Abe}(1940)}]{Abe:1940}%
  \BibitemOpen
  \bibfield  {author} {\bibinfo {author} {\bibfnamefont {M.}~\bibnamefont
  {Abe}},\ }\href {\doibase 10.4099/jjm1924.16.0_169} {\bibfield  {journal}
  {\bibinfo  {journal} {Jpn. J. Math.}\ }\textbf {\bibinfo {volume} {16}},\
  \bibinfo {pages} {169} (\bibinfo {year} {1940})}\BibitemShut {NoStop}%
\bibitem [{Note2()}]{Note2}%
  \BibitemOpen
  \bibinfo {note} {One can change this $\protect \boldsymbol {k}\cdot \protect
  \boldsymbol {p}$ model into a lattice model through the usual substitution
  $k_i \DOTSB \mapstochar \rightarrow \protect \qopname \relax o{sin}k_i$ and
  $k_i^2\DOTSB \mapstochar \rightarrow 2(1-\protect \qopname \relax
  o{cos}k_i)$, which may however result in many copies of the band nodes at
  other high-symmetry momenta. However, the non-Hermiticity of the system
  allows for an alternative substitution $k_i \DOTSB \mapstochar \rightarrow
  2\protect \qopname \relax o{sin}\protect \genfrac {}{}{}1{k_i}{2}\protect
  \qopname \relax o{exp}{(\protect \mathrm {i}k_i/2)}$ and $k_i^2\DOTSB
  \mapstochar \rightarrow 2(1-\protect \qopname \relax o{cos}k_i)$ that results
  in fewer band nodes elsewhere in the Brillouin zone.}\BibitemShut {Stop}%
\bibitem [{\citenamefont {Nielsen}\ and\ \citenamefont
  {Ninomiya}(1981)}]{Nielsen:1981}%
  \BibitemOpen
  \bibfield  {author} {\bibinfo {author} {\bibfnamefont {H.~B.}\ \bibnamefont
  {Nielsen}}\ and\ \bibinfo {author} {\bibfnamefont {M.}~\bibnamefont
  {Ninomiya}},\ }\href {\doibase 10.1016/0550-3213(81)90361-8} {\bibfield
  {journal} {\bibinfo  {journal} {Nucl. Phys.}\ }\textbf {\bibinfo {volume}
  {B185}},\ \bibinfo {pages} {20} (\bibinfo {year} {1981})}\BibitemShut
  {NoStop}%
\bibitem [{\citenamefont {Higashikawa}\ \emph {et~al.}(2019)\citenamefont
  {Higashikawa}, \citenamefont {Nakagawa},\ and\ \citenamefont
  {Ueda}}]{Higashikawa:2019}%
  \BibitemOpen
  \bibfield  {author} {\bibinfo {author} {\bibfnamefont {S.}~\bibnamefont
  {Higashikawa}}, \bibinfo {author} {\bibfnamefont {M.}~\bibnamefont
  {Nakagawa}}, \ and\ \bibinfo {author} {\bibfnamefont {M.}~\bibnamefont
  {Ueda}},\ }\href {\doibase 10.1103/PhysRevLett.123.066403} {\bibfield
  {journal} {\bibinfo  {journal} {Phys. Rev. Lett.}\ }\textbf {\bibinfo
  {volume} {123}},\ \bibinfo {pages} {066403} (\bibinfo {year}
  {2019})}\BibitemShut {NoStop}%
\bibitem [{\citenamefont {Yu}\ \emph {et~al.}(2019)\citenamefont {Yu},
  \citenamefont {Wu}, \citenamefont {Zhao},\ and\ \citenamefont
  {Yang}}]{Yu:2019}%
  \BibitemOpen
  \bibfield  {author} {\bibinfo {author} {\bibfnamefont {Z.-M.}\ \bibnamefont
  {Yu}}, \bibinfo {author} {\bibfnamefont {W.}~\bibnamefont {Wu}}, \bibinfo
  {author} {\bibfnamefont {Y.~X.}\ \bibnamefont {Zhao}}, \ and\ \bibinfo
  {author} {\bibfnamefont {S.~A.}\ \bibnamefont {Yang}},\ }\href {\doibase
  10.1103/PhysRevB.100.041118} {\bibfield  {journal} {\bibinfo  {journal}
  {Phys. Rev. B}\ }\textbf {\bibinfo {volume} {100}},\ \bibinfo {pages}
  {041118} (\bibinfo {year} {2019})}\BibitemShut {NoStop}%
\bibitem [{\citenamefont {Yang}\ \emph {et~al.}(2018)\citenamefont {Yang},
  \citenamefont {Guo}, \citenamefont {Tremain}, \citenamefont {Liu},
  \citenamefont {Barr}, \citenamefont {Yan}, \citenamefont {Gao}, \citenamefont
  {Liu}, \citenamefont {Xiang}, \citenamefont {Chen}, \citenamefont {Fang},
  \citenamefont {Hibbins}, \citenamefont {Lu},\ and\ \citenamefont
  {Zhang}}]{Yang:2018}%
  \BibitemOpen
  \bibfield  {author} {\bibinfo {author} {\bibfnamefont {B.}~\bibnamefont
  {Yang}}, \bibinfo {author} {\bibfnamefont {Q.}~\bibnamefont {Guo}}, \bibinfo
  {author} {\bibfnamefont {B.}~\bibnamefont {Tremain}}, \bibinfo {author}
  {\bibfnamefont {R.}~\bibnamefont {Liu}}, \bibinfo {author} {\bibfnamefont
  {L.~E.}\ \bibnamefont {Barr}}, \bibinfo {author} {\bibfnamefont
  {Q.}~\bibnamefont {Yan}}, \bibinfo {author} {\bibfnamefont {W.}~\bibnamefont
  {Gao}}, \bibinfo {author} {\bibfnamefont {H.}~\bibnamefont {Liu}}, \bibinfo
  {author} {\bibfnamefont {Y.}~\bibnamefont {Xiang}}, \bibinfo {author}
  {\bibfnamefont {J.}~\bibnamefont {Chen}}, \bibinfo {author} {\bibfnamefont
  {C.}~\bibnamefont {Fang}}, \bibinfo {author} {\bibfnamefont {A.}~\bibnamefont
  {Hibbins}}, \bibinfo {author} {\bibfnamefont {L.}~\bibnamefont {Lu}}, \ and\
  \bibinfo {author} {\bibfnamefont {S.}~\bibnamefont {Zhang}},\ }\href
  {\doibase 10.1126/science.aaq1221} {\bibfield  {journal} {\bibinfo  {journal}
  {Science}\ }\textbf {\bibinfo {volume} {359}},\ \bibinfo {pages} {1013}
  (\bibinfo {year} {2018})}\BibitemShut {NoStop}%
\bibitem [{\citenamefont {Guo}\ \emph {et~al.}(2019)\citenamefont {Guo},
  \citenamefont {You}, \citenamefont {Yang}, \citenamefont {Sellman},
  \citenamefont {Blythe}, \citenamefont {Liu}, \citenamefont {Xiang},
  \citenamefont {Li}, \citenamefont {Fan}, \citenamefont {Chen}, \citenamefont
  {Chan},\ and\ \citenamefont {Zhang}}]{Guo:2019}%
  \BibitemOpen
  \bibfield  {author} {\bibinfo {author} {\bibfnamefont {Q.}~\bibnamefont
  {Guo}}, \bibinfo {author} {\bibfnamefont {O.}~\bibnamefont {You}}, \bibinfo
  {author} {\bibfnamefont {B.}~\bibnamefont {Yang}}, \bibinfo {author}
  {\bibfnamefont {J.~B.}\ \bibnamefont {Sellman}}, \bibinfo {author}
  {\bibfnamefont {E.}~\bibnamefont {Blythe}}, \bibinfo {author} {\bibfnamefont
  {H.}~\bibnamefont {Liu}}, \bibinfo {author} {\bibfnamefont {Y.}~\bibnamefont
  {Xiang}}, \bibinfo {author} {\bibfnamefont {J.}~\bibnamefont {Li}}, \bibinfo
  {author} {\bibfnamefont {D.}~\bibnamefont {Fan}}, \bibinfo {author}
  {\bibfnamefont {J.}~\bibnamefont {Chen}}, \bibinfo {author} {\bibfnamefont
  {C.~T.}\ \bibnamefont {Chan}}, \ and\ \bibinfo {author} {\bibfnamefont
  {S.}~\bibnamefont {Zhang}},\ }\href {\doibase 10.1103/PhysRevLett.122.203903}
  {\bibfield  {journal} {\bibinfo  {journal} {Phys. Rev. Lett.}\ }\textbf
  {\bibinfo {volume} {122}},\ \bibinfo {pages} {203903} (\bibinfo {year}
  {2019})}\BibitemShut {NoStop}%
\bibitem [{\citenamefont {Yao}\ and\ \citenamefont {Wang}(2018)}]{Yao:2018a}%
  \BibitemOpen
  \bibfield  {author} {\bibinfo {author} {\bibfnamefont {S.}~\bibnamefont
  {Yao}}\ and\ \bibinfo {author} {\bibfnamefont {Z.}~\bibnamefont {Wang}},\
  }\href {\doibase 10.1103/PhysRevLett.121.086803} {\bibfield  {journal}
  {\bibinfo  {journal} {Phys. Rev. Lett.}\ }\textbf {\bibinfo {volume} {121}},\
  \bibinfo {pages} {086803} (\bibinfo {year} {2018})}\BibitemShut {NoStop}%
\bibitem [{\citenamefont {Yokomizo}\ and\ \citenamefont
  {Murakami}(2019)}]{Yokomizo:2019}%
  \BibitemOpen
  \bibfield  {author} {\bibinfo {author} {\bibfnamefont {K.}~\bibnamefont
  {Yokomizo}}\ and\ \bibinfo {author} {\bibfnamefont {S.}~\bibnamefont
  {Murakami}},\ }\href {\doibase 10.1103/PhysRevLett.123.066404} {\bibfield
  {journal} {\bibinfo  {journal} {Phys. Rev. Lett.}\ }\textbf {\bibinfo
  {volume} {123}},\ \bibinfo {pages} {066404} (\bibinfo {year}
  {2019})}\BibitemShut {NoStop}%
\bibitem [{\citenamefont {Lee}(2016)}]{Lee:2016}%
  \BibitemOpen
  \bibfield  {author} {\bibinfo {author} {\bibfnamefont {T.~E.}\ \bibnamefont
  {Lee}},\ }\href {\doibase 10.1103/PhysRevLett.116.133903} {\bibfield
  {journal} {\bibinfo  {journal} {Phys. Rev. Lett.}\ }\textbf {\bibinfo
  {volume} {116}},\ \bibinfo {pages} {133903} (\bibinfo {year}
  {2016})}\BibitemShut {NoStop}%
\bibitem [{\citenamefont {Song}\ \emph
  {et~al.}(2019{\natexlab{b}})\citenamefont {Song}, \citenamefont {Yao},\ and\
  \citenamefont {Wang}}]{Song:2019a}%
  \BibitemOpen
  \bibfield  {author} {\bibinfo {author} {\bibfnamefont {F.}~\bibnamefont
  {Song}}, \bibinfo {author} {\bibfnamefont {S.}~\bibnamefont {Yao}}, \ and\
  \bibinfo {author} {\bibfnamefont {Z.}~\bibnamefont {Wang}},\ }\href {\doibase
  10.1103/PhysRevLett.123.170401} {\bibfield  {journal} {\bibinfo  {journal}
  {Phys. Rev. Lett.}\ }\textbf {\bibinfo {volume} {123}},\ \bibinfo {pages}
  {170401} (\bibinfo {year} {2019}{\natexlab{b}})}\BibitemShut {NoStop}%
\bibitem [{\citenamefont {Kitaev}(2009)}]{Kitaev:2009}%
  \BibitemOpen
  \bibfield  {author} {\bibinfo {author} {\bibfnamefont {A.}~\bibnamefont
  {Kitaev}},\ }\href {\doibase 10.1063/1.3149495} {\bibfield  {journal}
  {\bibinfo  {journal} {AIP Conf. Proc.}\ }\textbf {\bibinfo {volume} {1134}},\
  \bibinfo {pages} {22} (\bibinfo {year} {2009})}\BibitemShut {NoStop}%
\bibitem [{\citenamefont {Hatcher}(2002)}]{Hatcher:2002}%
  \BibitemOpen
  \bibfield  {author} {\bibinfo {author} {\bibfnamefont {A.}~\bibnamefont
  {Hatcher}},\ }\href@noop {} {\emph {\bibinfo {title} {Algebraic Topology}}}\
  (\bibinfo  {publisher} {Cambridge University Press},\ \bibinfo {address}
  {Cambridge},\ \bibinfo {year} {2002})\BibitemShut {NoStop}%
\bibitem [{\citenamefont {Tiwari}\ and\ \citenamefont
  {Bzdu\ifmmode~\check{s}\else \v{s}\fi{}ek}(2020)}]{Tiwari:2019}%
  \BibitemOpen
  \bibfield  {author} {\bibinfo {author} {\bibfnamefont {A.}~\bibnamefont
  {Tiwari}}\ and\ \bibinfo {author} {\bibfnamefont {T.~c.~v.}\ \bibnamefont
  {Bzdu\ifmmode~\check{s}\else \v{s}\fi{}ek}},\ }\href {\doibase
  10.1103/PhysRevB.101.195130} {\bibfield  {journal} {\bibinfo  {journal}
  {Phys. Rev. B}\ }\textbf {\bibinfo {volume} {101}},\ \bibinfo {pages}
  {195130} (\bibinfo {year} {2020})}\BibitemShut {NoStop}%
\bibitem [{\citenamefont {Volovik}\ and\ \citenamefont
  {Mineev}(1977)}]{Volovik:1977}%
  \BibitemOpen
  \bibfield  {author} {\bibinfo {author} {\bibfnamefont {G.~E.}\ \bibnamefont
  {Volovik}}\ and\ \bibinfo {author} {\bibfnamefont {V.~P.}\ \bibnamefont
  {Mineev}},\ }\href {\doibase 10.1142/9789814317344_0018} {\bibfield
  {journal} {\bibinfo  {journal} {Zh. Eksp. Teor. Fiz}\ }\textbf {\bibinfo
  {volume} {72}},\ \bibinfo {pages} {2256} (\bibinfo {year}
  {1977})}\BibitemShut {NoStop}%
\bibitem [{\citenamefont {Alexander}\ \emph {et~al.}(2012)\citenamefont
  {Alexander}, \citenamefont {Chen}, \citenamefont {Matsumoto},\ and\
  \citenamefont {Kamien}}]{Alexander:2012}%
  \BibitemOpen
  \bibfield  {author} {\bibinfo {author} {\bibfnamefont {G.~P.}\ \bibnamefont
  {Alexander}}, \bibinfo {author} {\bibfnamefont {B.~G.-g.}\ \bibnamefont
  {Chen}}, \bibinfo {author} {\bibfnamefont {E.~A.}\ \bibnamefont {Matsumoto}},
  \ and\ \bibinfo {author} {\bibfnamefont {R.~D.}\ \bibnamefont {Kamien}},\
  }\href {\doibase 10.1103/RevModPhys.84.497} {\bibfield  {journal} {\bibinfo
  {journal} {Rev. Mod. Phys.}\ }\textbf {\bibinfo {volume} {84}},\ \bibinfo
  {pages} {497} (\bibinfo {year} {2012})}\BibitemShut {NoStop}%
\bibitem [{\citenamefont {Arkowitz}(2011)}]{Arkowitz:2011}%
  \BibitemOpen
  \bibfield  {author} {\bibinfo {author} {\bibfnamefont {M.}~\bibnamefont
  {Arkowitz}},\ }\href@noop {} {\emph {\bibinfo {title} {Introduction to
  homotopy theory}}}\ (\bibinfo  {publisher} {Springer Science \& Business
  Media},\ \bibinfo {year} {2011})\BibitemShut {NoStop}%
\bibitem [{Note3()}]{Note3}%
  \BibitemOpen
  \bibinfo {note} {The compatibility further requires $\triangleright _g
  \protect \tmspace -\thinmuskip {.1667em}\circ \protect \tmspace -\thinmuskip
  {.1667em} \triangleright _h \protect \tmspace -\thinmuskip {.1667em}=\protect
  \tmspace -\thinmuskip {.1667em} \triangleright _{g\circ h}$. The collection
  $[\pi _1(M), \pi _2(M), \triangleright ]$, together with an additional piece
  of data called the Postnikov class, form a mathematical structure called the
  \protect \emph {fundamental 2-group of $M$}~\cite
  {Baez:2004,Ang:2018rls}.}\BibitemShut {Stop}%
\bibitem [{Note4()}]{Note4}%
  \BibitemOpen
  \bibinfo {note} {Our follow-up work in Ref.~\cite {Wojcik:2019} adopts a
  different strategy, and expresses the same space as $M = S^2 \times S^1
  /\protect \mathbbold {Z}_2$, where the quotient identifies antipodal points
  $(x,y)\sim (-x,-y)$ in $S^2\times S^1$.}\BibitemShut {Stop}%
\bibitem [{Note5()}]{Note5}%
  \BibitemOpen
  \bibinfo {note} {This follows from the long exact sequence of relative
  homotopy groups for pair $(\protect \mathsf {G},\protect \mathsf {H})$~\cite
  {Sun:2018,Hatcher:2002}.}\BibitemShut {Stop}%
\bibitem [{\citenamefont {Brody}(2013)}]{Brody:2013}%
  \BibitemOpen
  \bibfield  {author} {\bibinfo {author} {\bibfnamefont {D.~C.}\ \bibnamefont
  {Brody}},\ }\href {\doibase 10.1088/1751-8113/47/3/035305} {\bibfield
  {journal} {\bibinfo  {journal} {J. Phys. A}\ }\textbf {\bibinfo {volume}
  {47}},\ \bibinfo {pages} {035305} (\bibinfo {year} {2013})}\BibitemShut
  {NoStop}%
\bibitem [{\citenamefont {Kunst}\ \emph {et~al.}(2018)\citenamefont {Kunst},
  \citenamefont {Edvardsson}, \citenamefont {Budich},\ and\ \citenamefont
  {Bergholtz}}]{Kunst:2018}%
  \BibitemOpen
  \bibfield  {author} {\bibinfo {author} {\bibfnamefont {F.~K.}\ \bibnamefont
  {Kunst}}, \bibinfo {author} {\bibfnamefont {E.}~\bibnamefont {Edvardsson}},
  \bibinfo {author} {\bibfnamefont {J.~C.}\ \bibnamefont {Budich}}, \ and\
  \bibinfo {author} {\bibfnamefont {E.~J.}\ \bibnamefont {Bergholtz}},\ }\href
  {\doibase 10.1103/PhysRevLett.121.026808} {\bibfield  {journal} {\bibinfo
  {journal} {Phys. Rev. Lett.}\ }\textbf {\bibinfo {volume} {121}},\ \bibinfo
  {pages} {026808} (\bibinfo {year} {2018})}\BibitemShut {NoStop}%
\bibitem [{\citenamefont {Martinez~Alvarez}\ \emph {et~al.}(2018)\citenamefont
  {Martinez~Alvarez}, \citenamefont {Barrios~Vargas}, \citenamefont
  {Berdakin},\ and\ \citenamefont {Foa~Torres}}]{Martinez:2018}%
  \BibitemOpen
  \bibfield  {author} {\bibinfo {author} {\bibfnamefont {V.~M.}\ \bibnamefont
  {Martinez~Alvarez}}, \bibinfo {author} {\bibfnamefont {J.~E.}\ \bibnamefont
  {Barrios~Vargas}}, \bibinfo {author} {\bibfnamefont {M.}~\bibnamefont
  {Berdakin}}, \ and\ \bibinfo {author} {\bibfnamefont {L.~E.~F.}\ \bibnamefont
  {Foa~Torres}},\ }\href {\doibase 10.1140/epjst/e2018-800091-5} {\bibfield
  {journal} {\bibinfo  {journal} {Eur. Phys. J. Spec. Top.}\ }\textbf {\bibinfo
  {volume} {227}},\ \bibinfo {pages} {1295} (\bibinfo {year}
  {2018})}\BibitemShut {NoStop}%
\bibitem [{\citenamefont {Procesi}(2003)}]{Procesi:2007}%
  \BibitemOpen
  \bibfield  {author} {\bibinfo {author} {\bibfnamefont {C.}~\bibnamefont
  {Procesi}},\ }\href@noop {} {\emph {\bibinfo {title} {Lie {G}roups: {A}n
  {A}pproach through {I}nvariants and {R}epresentations}}}\ (\bibinfo
  {publisher} {Springer},\ \bibinfo {address} {New York},\ \bibinfo {year}
  {2003})\BibitemShut {NoStop}%
\bibitem [{\citenamefont {Ahn}\ \emph {et~al.}(2018{\natexlab{a}})\citenamefont
  {Ahn}, \citenamefont {Kim}, \citenamefont {Youngkuk},\ and\ \citenamefont
  {Yang}}]{Ahn:2018}%
  \BibitemOpen
  \bibfield  {author} {\bibinfo {author} {\bibfnamefont {J.}~\bibnamefont
  {Ahn}}, \bibinfo {author} {\bibfnamefont {D.}~\bibnamefont {Kim}}, \bibinfo
  {author} {\bibfnamefont {K.}~\bibnamefont {Youngkuk}}, \ and\ \bibinfo
  {author} {\bibfnamefont {B.-J.~Y.}\ \bibnamefont {Yang}},\ }\href {\doibase
  10.1103/PhysRevLett.121.106403} {\bibfield  {journal} {\bibinfo  {journal}
  {Phys. Rev. Lett.}\ }\textbf {\bibinfo {volume} {121}},\ \bibinfo {pages}
  {106403} (\bibinfo {year} {2018}{\natexlab{a}})}\BibitemShut {NoStop}%
\bibitem [{\citenamefont {Wu}\ \emph {et~al.}(2019)\citenamefont {Wu},
  \citenamefont {Soluyanov},\ and\ \citenamefont {Bzdu\v{s}ek}}]{Wu:2018b}%
  \BibitemOpen
  \bibfield  {author} {\bibinfo {author} {\bibfnamefont {Q.}~\bibnamefont
  {Wu}}, \bibinfo {author} {\bibfnamefont {A.~A.}\ \bibnamefont {Soluyanov}}, \
  and\ \bibinfo {author} {\bibfnamefont {T.}~\bibnamefont {Bzdu\v{s}ek}},\
  }\href {https://dx.doi.org/10.1126/science.aau8740} {\bibfield  {journal}
  {\bibinfo  {journal} {Science}\ }\textbf {\bibinfo {volume} {365}},\ \bibinfo
  {pages} {1273} (\bibinfo {year} {2019})}\BibitemShut {NoStop}%
\bibitem [{\citenamefont {Ahn}\ \emph {et~al.}(2018{\natexlab{b}})\citenamefont
  {Ahn}, \citenamefont {Park},\ and\ \citenamefont {Yang}}]{Ahn:2018b}%
  \BibitemOpen
  \bibfield  {author} {\bibinfo {author} {\bibfnamefont {J.}~\bibnamefont
  {Ahn}}, \bibinfo {author} {\bibfnamefont {S.}~\bibnamefont {Park}}, \ and\
  \bibinfo {author} {\bibfnamefont {B.-J.}\ \bibnamefont {Yang}},\ }\href@noop
  {} {\bibfield  {journal} {\bibinfo  {journal} {ArXiv e-prints}\ } (\bibinfo
  {year} {2018}{\natexlab{b}})},\ \Eprint {http://arxiv.org/abs/1808.05375}
  {arXiv:1808.05375} \BibitemShut {NoStop}%
\bibitem [{\citenamefont {Bouhon}\ \emph {et~al.}(2019)\citenamefont {Bouhon},
  \citenamefont {Wu}, \citenamefont {Slager}, \citenamefont {Weng},
  \citenamefont {Yazyev},\ and\ \citenamefont {Bzdu\v{s}ek}}]{Bouhon:2019}%
  \BibitemOpen
  \bibfield  {author} {\bibinfo {author} {\bibfnamefont {A.}~\bibnamefont
  {Bouhon}}, \bibinfo {author} {\bibfnamefont {Q.}~\bibnamefont {Wu}}, \bibinfo
  {author} {\bibfnamefont {R.-J.}\ \bibnamefont {Slager}}, \bibinfo {author}
  {\bibfnamefont {H.}~\bibnamefont {Weng}}, \bibinfo {author} {\bibfnamefont
  {O.~V.}\ \bibnamefont {Yazyev}}, \ and\ \bibinfo {author} {\bibfnamefont
  {T.}~\bibnamefont {Bzdu\v{s}ek}},\ }\href
  {https://arxiv.org/abs/1907.10611v3} {\bibfield  {journal} {\bibinfo
  {journal} {arXiv:1907.10611}\ } (\bibinfo {year} {2019})}\BibitemShut
  {NoStop}%
\bibitem [{\citenamefont {Baez}\ and\ \citenamefont {Lauda}(2004)}]{Baez:2004}%
  \BibitemOpen
  \bibfield  {author} {\bibinfo {author} {\bibfnamefont {J.~C.}\ \bibnamefont
  {Baez}}\ and\ \bibinfo {author} {\bibfnamefont {A.~D.}\ \bibnamefont
  {Lauda}},\ }\href@noop {} {\bibfield  {journal} {\bibinfo  {journal} {Theory
  Appl. Categ.}\ }\textbf {\bibinfo {volume} {12}},\ \bibinfo {pages} {423}
  (\bibinfo {year} {2004})}\BibitemShut {NoStop}%
\bibitem [{\citenamefont {Ang}\ and\ \citenamefont
  {Prakash}(2018)}]{Ang:2018rls}%
  \BibitemOpen
  \bibfield  {author} {\bibinfo {author} {\bibfnamefont {J.~P.}\ \bibnamefont
  {Ang}}\ and\ \bibinfo {author} {\bibfnamefont {A.}~\bibnamefont {Prakash}},\
  }\href {https://arxiv.org/abs/1810.12965} {\bibfield  {journal} {\bibinfo
  {journal} {arXiv:1810.12965}\ } (\bibinfo {year} {2018})}\BibitemShut
  {NoStop}%
\end{thebibliography}%
\bibliographystyle{apsrev4-1}  
\end{document}